\documentclass{article} 
\usepackage[preprint]{colm2026_conference}

\usepackage{microtype}
\usepackage{hyperref}
\usepackage{url}
\usepackage{booktabs}
\usepackage{lineno}
\usepackage{graphicx}
\usepackage{subcaption}
\usepackage{xcolor}
\usepackage{amsmath}
\usepackage{tabularx}
\usepackage{multirow}
\usepackage{tcolorbox}
\usepackage{caption}
\tcbuselibrary{breakable}

\newtcolorbox{promptbox}{
  colback=gray!5,
  colframe=black!70,
  boxrule=0.5pt,
  arc=2pt,
  left=6pt,
  right=6pt,
  top=6pt,
  bottom=6pt,
  before upper=\setlength{\parskip}{0.8\baselineskip}
}

\definecolor{darkblue}{rgb}{0, 0, 0.5}
\hypersetup{colorlinks=true, citecolor=darkblue, linkcolor=darkblue, urlcolor=darkblue}

\newcommand{\dang}[1]{\textcolor{red}{DN: #1}}
\newcommand{\ari}[1]{\textcolor{magenta}{#1 --AH}}
\newcommand{\chenhao}[1]{\textcolor{blue}{#1 --CT}}

\title{Moral Mazes in the Era of LLMs}

\author{Dang Nguyen \\
University of Chicago \\
\texttt{dangnguyen@uchicago.edu} \\
\And
Harvey Yiyun Fu \\
University of Chicago \\
\texttt{harveyfu@uchicago.edu} \\
\And
Peter West \\
University of British Columbia \\
\texttt{pwest@cs.ubc.ca} \\
\And
Ari Holtzman \\
University of Chicago \\
\texttt{aholtzman@uchicago.edu} \\
\And
Chenhao Tan \\
University of Chicago \\
\texttt{chenhao@uchicago.edu} \\
}

\newcommand{\hrsim}[0]{HR Simulator\texttrademark}

\renewcommand{\dang}[1]{}
\renewcommand{\chenhao}[1]{}
\renewcommand{\ari}[1]{}

\begin{document}

\ifcolmsubmission
\linenumbers
\fi

\maketitle

\begin{abstract}

Navigating complex social situations is an integral part of corporate life,
ranging from giving critical feedback without hurting morale to rejecting requests without alienating teammates.
Although large language models (LLMs) are permeating the workplace, it is unclear how well they can navigate these norms.
To investigate this question, we created HR Simulator, a game where users roleplay as an HR officer and write emails to tackle challenging workplace scenarios, evaluated with GPT-4o as a judge based on scenario-specific rubrics.
We analyze over 600 human and LLM emails and find systematic differences in style: LLM emails are more formal and empathetic.
Furthermore, humans underperform LLMs (e.g., 23.5\% vs.\ 48--54\% scenario pass rate), but human emails rewritten by LLMs can outperform both, which indicates a hybrid advantage.
On the evaluation side, judges can exhibit differences in their email preferences: an analysis of 10 judge models reveals evidence for \emph{emergent tact}, where weaker models prefer direct, blunt communication but stronger models prefer more subtle messages.
Judges also agree with each other more as they scale, which hints at a convergence toward shared communicative norms that may differ from humans’.
Overall, our results suggest LLMs could substantially reshape communication in the workplace if they are widely adopted in professional correspondence.
\end{abstract}

\section{Introduction}

In \textit{Moral Mazes}, \citet{jackall1988moral} documented the unwritten rules of communication that govern corporate life.
Managers learn, often through painful experience, how to navigate a landscape where what you say matters less than how and when you say it.
A manager who discovers a serious problem cannot simply report it up the chain---doing so risks making their boss look negligent for not catching it first.
Instead, they learn to ``float'' the information sideways through informal channels, so that by the time a decision is needed, the right people know but no one is on record as the bearer of bad news.
This indirection is not a failure of communication; it \textit{is} the communication.
Being too direct, too blunt, or too transparent can be a career-ending move.

Large language models (LLMs) are about to enter this landscape.
They are already being used to draft and rephrase delicate business communications~\citep{chatterji2025people}, and integrations like Gemini in Gmail and autonomous assistants like OpenClaw~\citep{barnes2026gmail, superhuman2026, poke2026, steinberger2026openclaw} point to a near future where LLMs read, write, and respond to emails on our behalf.
But in contrast to tasks like mathematics where outputs are objective and verifiable, corporate communication is nuanced and often intentionally ambiguous.
How will LLMs navigate norms that took humans decades of organizational life to develop?
Will they adopt existing conventions, or will their own tendencies reshape how organizations communicate?
When a situation calls for strategic ambiguity, will they be too literal?
And if they develop their own communicative preferences, will those preferences align with ours~\citep{cloud2025subliminal, wu2024generative}?

To unravel these questions, we designed \hrsim
, a game where players take the role of a Human Resources officer and write emails to solve interpersonal issues at a fictional company.
It features a range of socially challenging scenarios. 
A successful message must thread the needle between various contrasting objectives, such as being warm versus distant, subtle versus direct.
We use GPT-4o (2024-11-20) to simulate characters who receive and respond to the emails (indicating whether they pass a scenario), which allows players to experience the office of the future where some emails will be read solely by LLMs.

\begin{figure}[t]
  \centering
  \begin{subfigure}{0.465\linewidth}
    \centering
    \includegraphics[width=\linewidth]{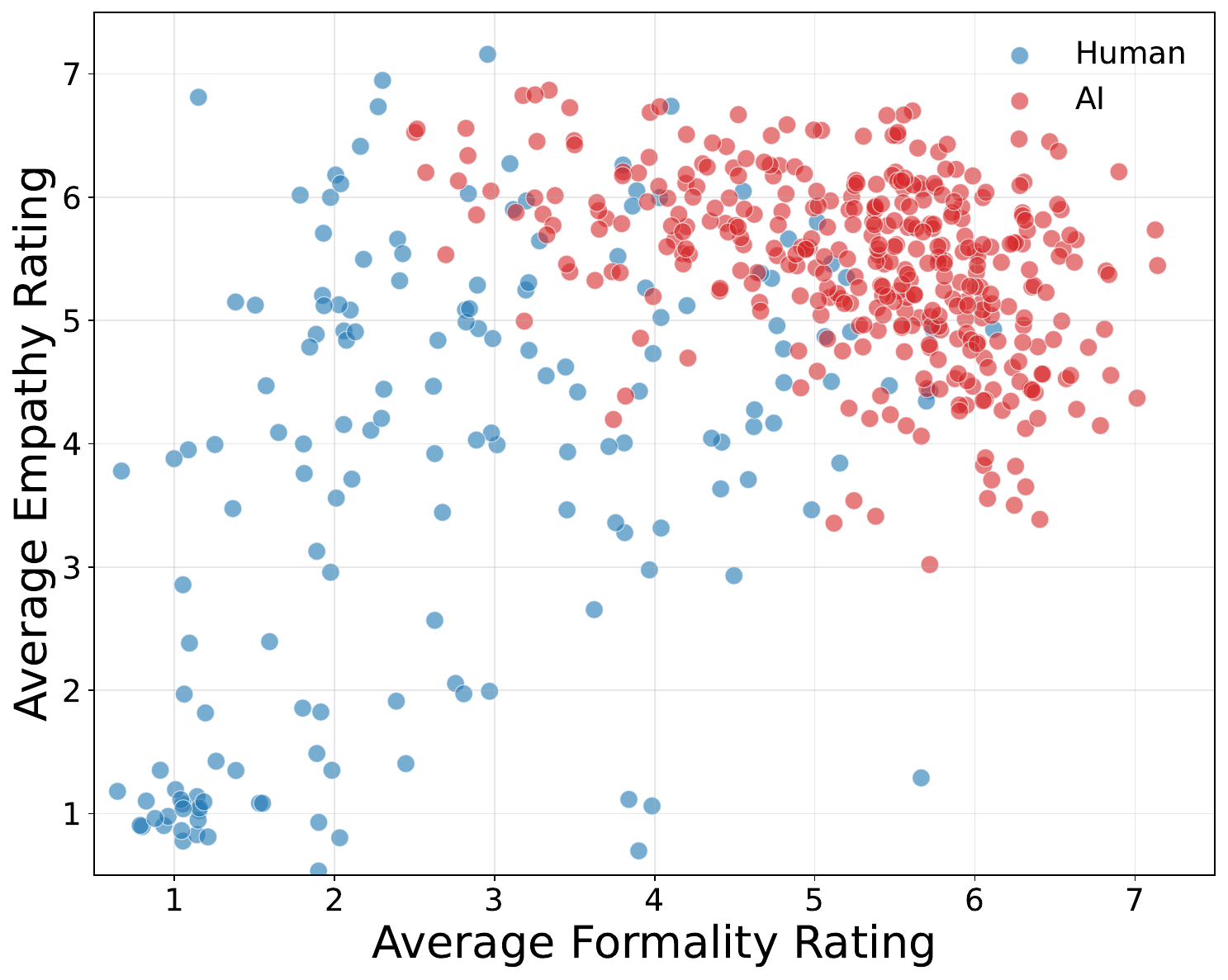}
    \caption{Human vs LLM emails.}
    \label{fig:human_llm_emp_form}
  \end{subfigure}\hfill
  \begin{subfigure}{0.445\linewidth}
    \centering
    \includegraphics[width=\linewidth]{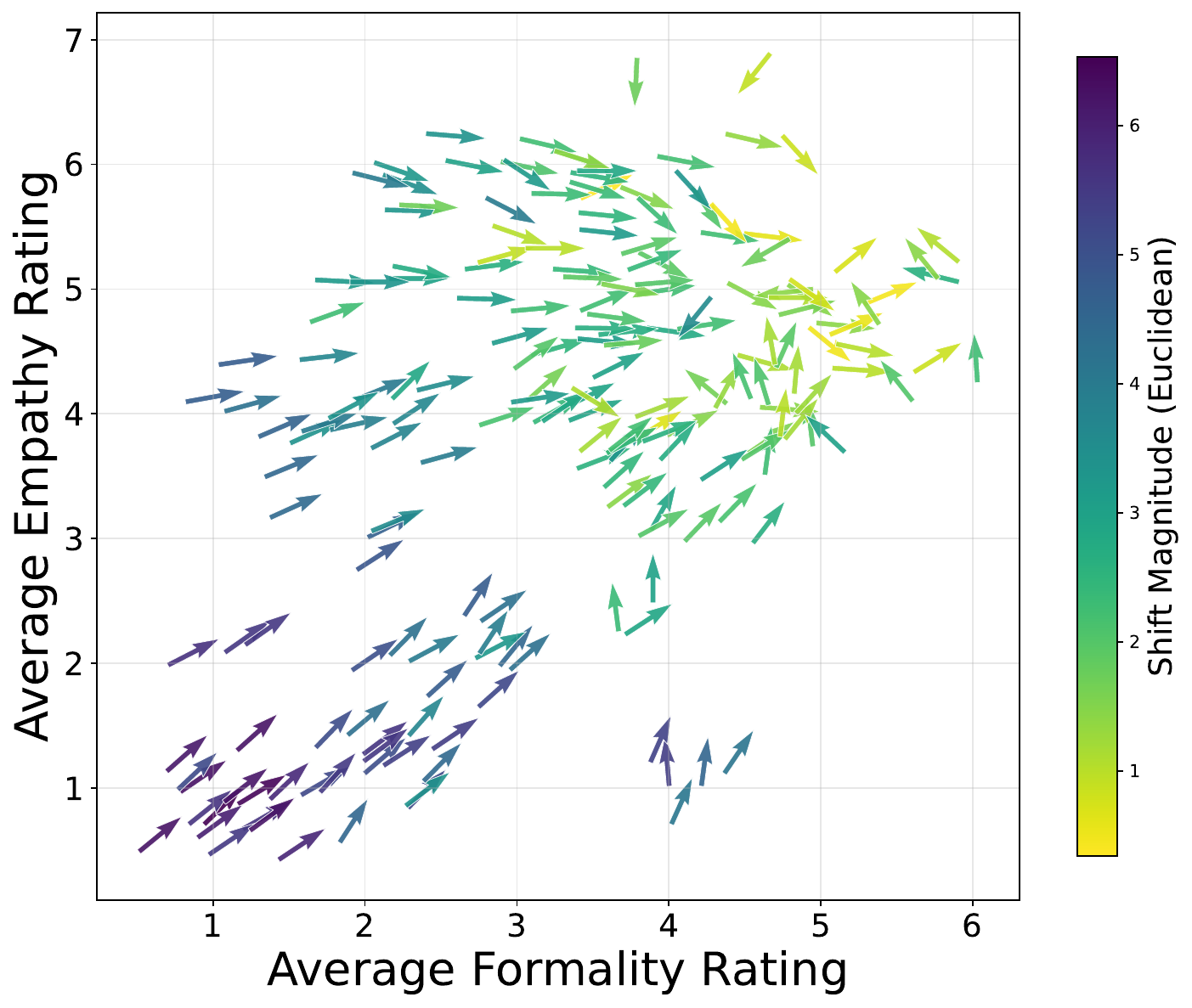}
    \caption{Where LLM rewrites take human emails.}
    \label{fig:human_llm_vectors}
  \end{subfigure}
  \caption{LLMs tend to write high empathy, high formality emails, whereas human emails are more diverse. When rewriting, LLMs take human emails toward the high-empathy high-formality quadrant. 
  We validate the scores with human annotations and find good agreement between the LLM annotator and humans, detailed in Appendix~\ref{appendix:tone_analysis}.
  }
  \label{fig:emp_form_analysis}
\end{figure}

We analyze emails' tone along two dimensions, \textbf{empathy} and \textbf{formality}, which capture the core tensions in the scenarios between how much one accommodates the recipient and whether one writes casually or professionally.
An analysis of over 600 human and LLM emails reveals systematic differences in style: LLM emails are predominantly formal and empathetic, whereas human emails are more varied (Figure~\ref{fig:emp_form_analysis}).
LLM judges consistently rate LLM-written emails more highly than human-written ones.
However, when LLMs rewrite human emails, they not only make the emails more formal and empathetic, but the resulting human+LLM emails can outperform \textit{both} human and LLM only, suggesting a hybrid advantage to email-writing.


Beyond GPT-4o, we evaluate emails with 10 judge models from various families and sizes and find not all judges agree on what makes a good email.
The evidence points toward \textit{emergent tact}---operationalized as whether an email navigates a social situation without over-communicating or creating future pitfalls---smaller judges prefer more direct, less tactful emails while larger judges prefer more subtle, tactful ones.
As models scale up, their email judgments increasingly agree, with the most advanced reaching about 0.5 Krippendorff's $\alpha$.
Combined with their ubiquitous adoption, this convergence can reshape the norms of workplace communication in ways that may diverge from human judgment. Our study makes a first step toward charting out this novel landscape of corporate ethics.

To summarize, we make the following contributions:
\begin{itemize}
    \item We create \hrsim~to study the emerging landscape of corporate communication norms shaped by LLMs.
    \item We analyze emails' tone and find LLMs tend to write more formal and empathetic emails, while human emails are more varied.
    \item Under LLM judges, human emails underpeform LLM emails, but a \textit{hybrid advantage} over both can be found in some human+LLM emails.
    \item Weaker judges prefer more direct emails, while stronger judges prefer more subtle emails, a phenomenon we term \textit{emergent tact}.
    \item LLMs increasingly agree on email judgments as they scale up, which can potentially reshape corporate communication norms.
    
    
    
\end{itemize}

\section{\hrsim: Measuring Communication In-Context}
\label{sec:hrsim}

\begin{figure}[t]
    \centering
    \includegraphics[width=0.90\textwidth]{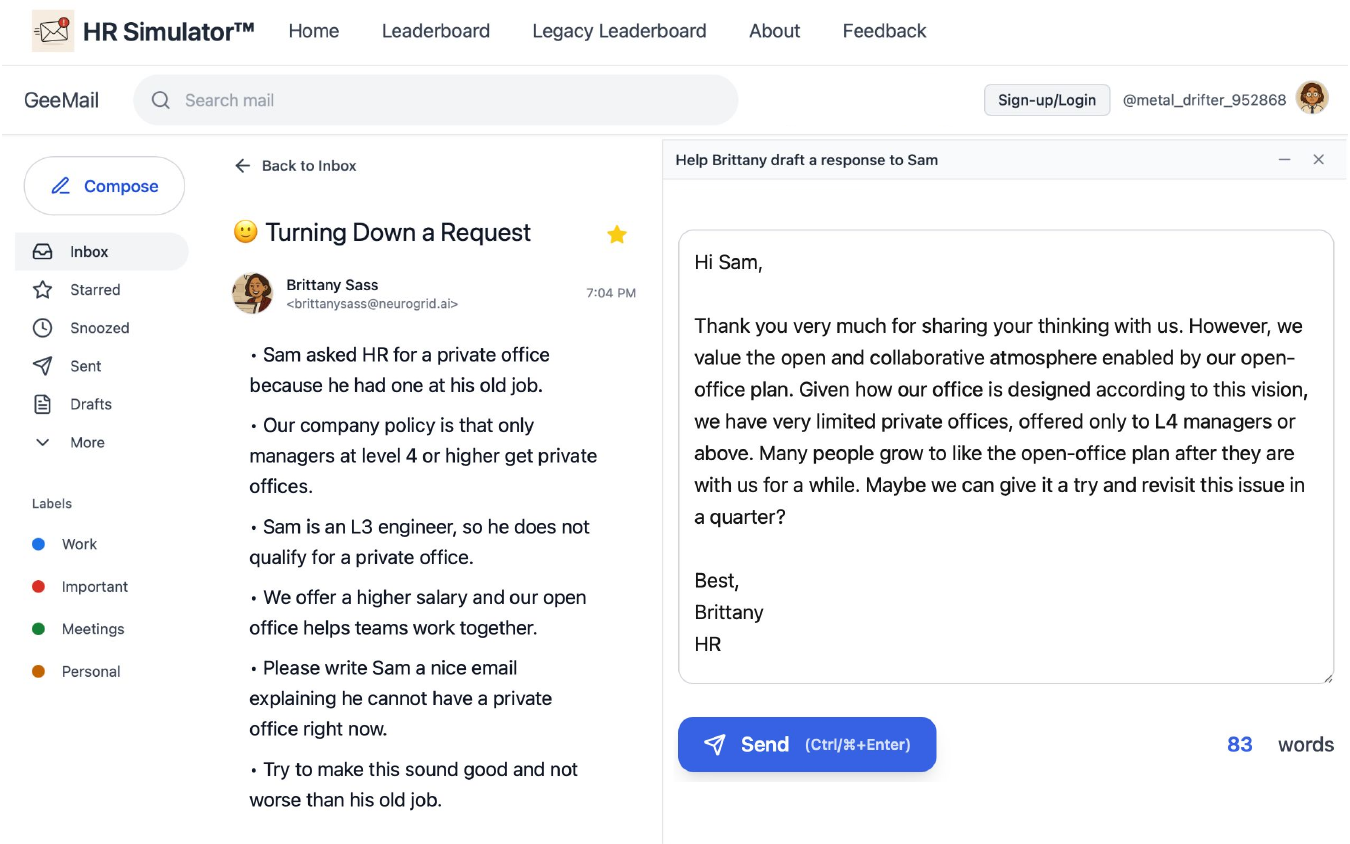}
    \caption{\hrsim~game interface. More details can be found in Appendix~\ref{appendix:hr_simulator}.}
    \label{fig:hr_sim_email_view}
\end{figure}

\hrsim~is a game where communication is the key mechanic.
The player plays as an HR officer and writes emails to solve interpersonal issues at a fictional company called ``NeuroGrid.''
There are five scenarios in total, unordered, each carefully designed to incorporate social tensions and trade-offs between different communication choices.
The player passes a scenario by sending an email that achieves the goal as evaluated by GPT-4o.
The game setting, story, and characters create incentives to meaningfully measure communication skills:
\textbf{Scenario 1---Declining a request:}
A newly hired, valuable employee joins NeuroGrid as a software engineer.
He asks for a private office, which is only available for L4 managers and above.
The player must decline the request without dampening his enthusiasm for the role.
\textbf{Scenario 4---Information seeking:}
A senior systems engineer is teamed with younger machine learning engineers on a project.
The former prefers a methodical approach while the latter prefer fast iteration.
The senior engineer is dissatisfied with this dynamic but is unaware of the true reason.
The player must talk to him to figure out the problem and propose a solution.
Full details on the scenarios and evaluation pipeline are in Appendix~\ref{appendix:hr_simulator}.

\begin{figure}[t]
    \centering
    \includegraphics[width=0.95\linewidth]{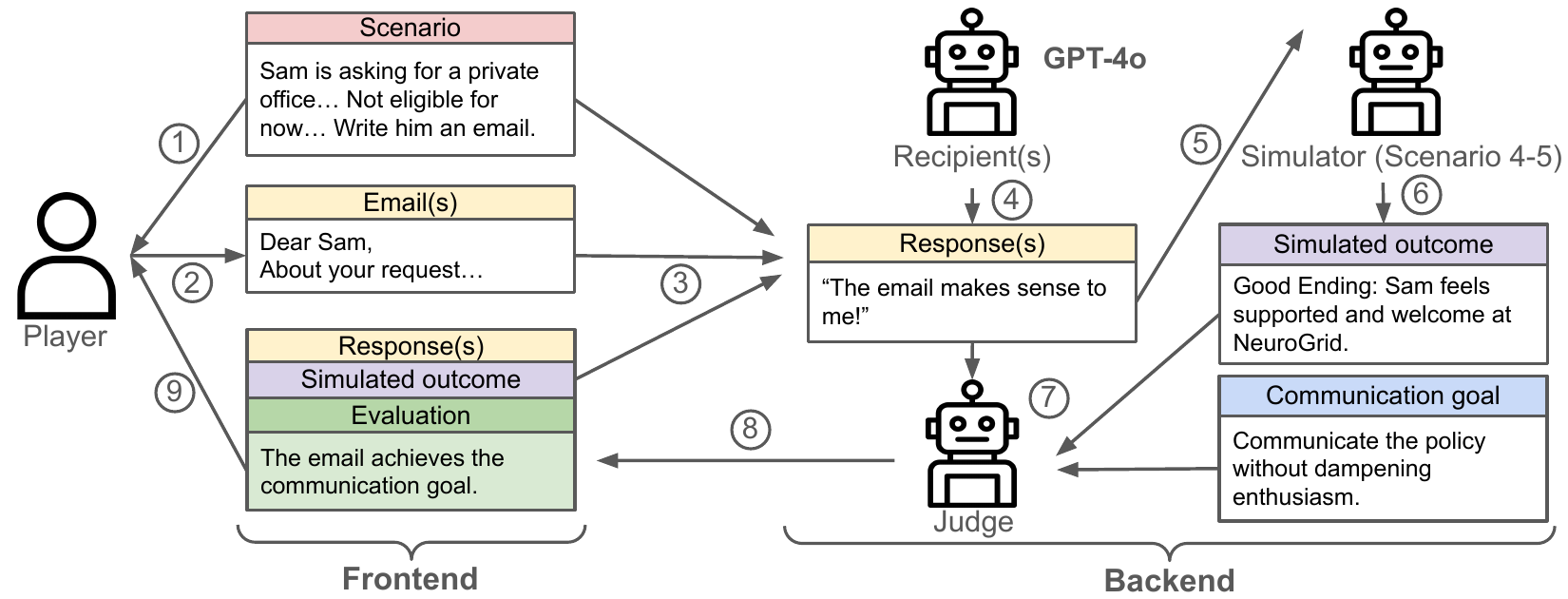}
    \caption{
    HR Simulator system. 
    The player reads a scenario email and responds.
    The scenario context and player's email are sent to the backend.
    The Recipient(s) responds to the player's email in-character.
    In scenarios 4 and 5, the Simulator simulates an outcome.
    The Judge produces an evaluation which is sent to the frontend.
    The user reads the results and decide whether to re-attempt.
    }
    \label{fig:hr_sim_system}
\end{figure}

An overview of the pipeline is provided in Figure~\ref{fig:hr_sim_system}.
On the frontend, the player initially sees the scenario description in the form of a task email as in Figure~\ref{fig:hr_sim_email_view}.
The UI parodies Gmail's UI to create an intuitive interface.
The user can click ``Reply,'' draft a response to the scenario, and send it to the backend, where the Recipient model reads the information and responds in character.
Different scenarios vary in configuration, such as the number of recipients (scenario 2) or whether the Simulator generates future events to surface an email's long-term effects (scenarios 4 and 5).
Then all information---scenario context, exchanged emails, optional simulated outcome, communication goal---are sent to the Judge for a final evaluation and then displayed to the frontend.
More details can be found in Appendix~\ref{appendix:hr_simulator}. 


\section{Methods}
\label{sec:method}

\paragraph{Writer and reader models.}
We analyze \hrsim~data from both the email writers' and readers' perspectives.
We examine different ways emails are written: entirely by humans, entirely by LLMs, or by a combination of both, where human emails are edited by LLMs to improve their quality.
We use a wide range of LLM writers, covering smaller, faster models such as Gemini 2.5 Flash and larger, more capable models such as Kimi K2 (Figure~\ref{fig:hybrid_advantage}).
Since future workplace environments will likely employ diverse models as email readers, we performed post-hoc analyses examining how different LLM judges rate emails.
The judge models also span a range of capability, including relatively \textbf{weaker} models such as GPT-4o 2024-11-20 (rank 127 on LM Arena),
and \textbf{stronger} models such as
Gemini 3 Flash (rank 3).
The main metric for comparing writers' performance on \hrsim~is the \textbf{pass rate}: the percentage of email attempts that pass a scenario.

\paragraph{Eliciting judges' email preferences.}
Among emails with the same outcome, a judge might prefer some over others.
To explore LLMs' preferences more finely, we examine each judge model's \textbf{ranking} of emails by pairwise comparing emails in the same scenario and fitting an Elo model to the outcomes~\citep{elo1978rating}.
Given a scenario, each email is compared against emails from a randomly sampled subset.
The judge picks the email that better achieves the communication goal.
We then compute Elo scores with the formula $E_A = \frac{1}{1 + 10^{(R_B - R_A)/400}}$ and update $R_A \leftarrow R_A + K(S_A - E_A)$~\citep{elo1978rating}.
$R_A$ and $R_B$ are the current ratings of emails $A$ and $B$, and $E_A$ is the predicted probability that $A$ will be preferred over $B$.
The actual outcome $S_A$ is 1 if email $A$ is chosen and 0 otherwise.
$K$ is a sensitivity constant that determines the magnitude of the rating update.
Initially, all emails are assigned a baseline rating (e.g., 1500).
Higher Elo scores indicate that an email is more frequently preferred by the judge relative to other emails in that scenario.
Full details can be found in Appendix~\ref{appendix:emergent_tact}. 

\paragraph{Metrics to charaterize emails.}
Beyond pass rates, to understand the stylistic and strategic differences between human and LLM emails, we annotate emails with Gemini 3 Flash on empathy, formality, and tact.
Empathy refers to how much an email attempts to understand and help the recipient with their concern.
Formality refers to whether an email's tone is casual or corporate.
While empathy and formality describe an email's \textit{tone}, tact captures the \textit{communication strategies}, referring to \textit{what} an email says to navigate a social situation: whether it delivers the message without over-communicating, which can lead to future pitfalls.
For example, in scenario 1, when declining the private office request, tactful emails focus on the appeal of the open office plan, while less tactful ones attempt to accomodate with more superficial perks like noise-canceling headphones.
Since tact is context-dependent, we tailor its definition to each scenario (full definitions in Appendix~\ref{appendix:emergent_tact}, Tables~\ref{tab:tactfulness_lvl_1_def}--\ref{tab:tactfulness_lvl_5_def}).
For all three descriptors, we use Gemini 3 Flash to annotate each paragraph on a Likert scale of 1 to 7 and average the scores per email.
\section{Humans and LLMs Communicate Differently}
\label{sec:results}


\subsection{Differences in Email Tone}

\paragraph{LLMs write formal and empathetic emails while humans write more diverse ones.}
Figure~\ref{fig:human_llm_emp_form} shows LLM emails are predominantly formal and empathetic.
In contrast, human emails are more diffuse, also occupying
the top and bottom left quadrants.
Few emails naturally fall into the high formality low empathy quadrant (bottom right).
For LLMs, this could be because they do not naturally write low empathy emails \textit{in general}, whereas for human players,
they might lack the skills to be both uncompromising and professional.
Human emails that are low empathy are also low formality because their content is often curt, dismissive, and sometimes conveys an annoyance that the player felt upon repeated failures.
Details on annotations and examples of emails can be found in Appendix~\ref{appendix:tone_analysis}.


\paragraph{LLM rewrites take human emails toward the high-formality high-empathy quadrant.}
We examine what happens to a human email's empathy and formality ratings when it is rewritten by an LLM.
Figure~\ref{fig:human_llm_vectors} shows that LLM rewrites take human emails toward the top right quadrant, i.e., toward the natural region of LLM emails.
The vectors' colors indicate their magnitude: emails further away from the top right quadrant are rewritten more drastically to land inside the quadrant.
This suggests that an email's tone plays a role in determining its outcome.
Many human emails contain the right message but are conveyed in a wrong tone, and rewriting for more professionalism likely helped flip the decision.
Examples of how LLMs rewrite human emails to the high-formality high-empathy quadrant can be found in Appendix~\ref{appendix:tone_analysis}.

\subsection{The Human+LLM Hybrid Advantage in Communication}

\begin{figure}[t]
\centering \includegraphics[width=0.9\linewidth]{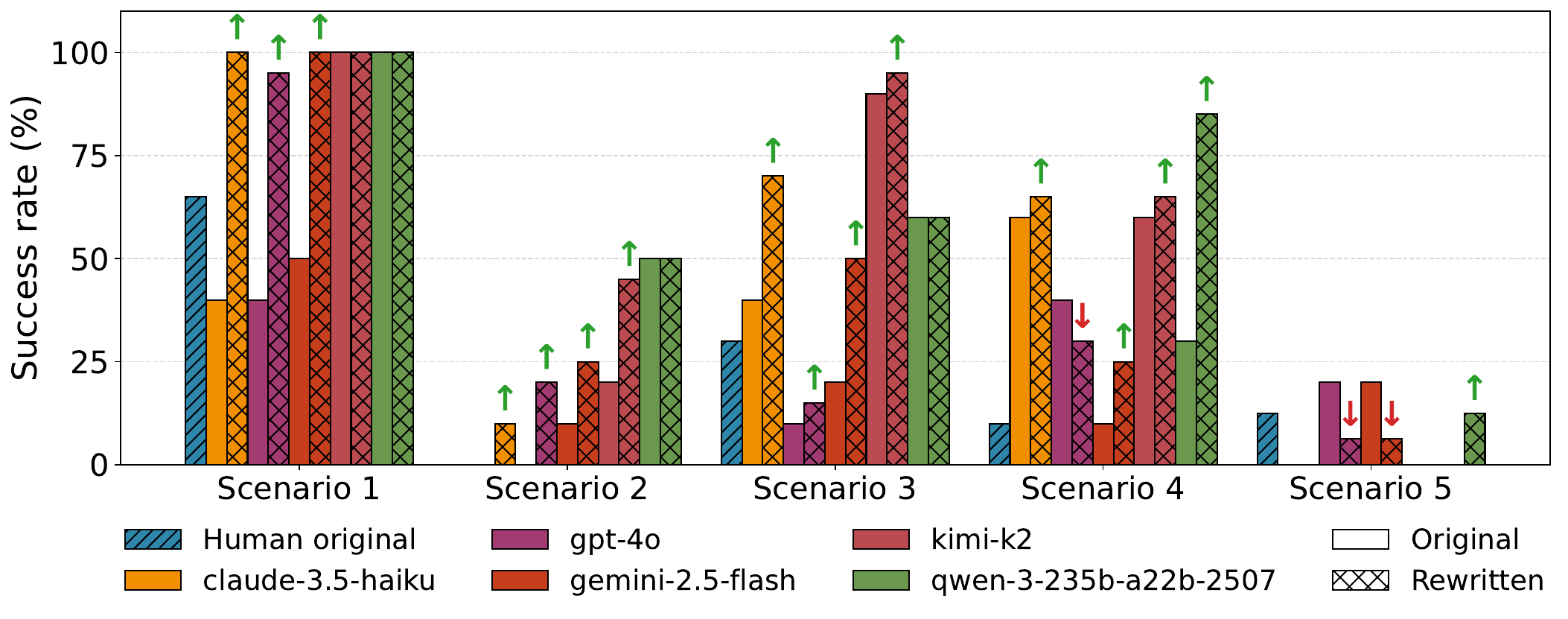}

\caption{The hybrid advantage.
Green arrows denote when the Human+LLM pass rate is higher than that of LLM-only, while red arrows denote when it is lower.
}
\label{fig:hybrid_advantage}
\end{figure}

\paragraph{Humans underperform the best LLMs at email communication.}
Figure~\ref{fig:hybrid_advantage} shows the performance of human and AI players on \hrsim~as judged by GPT-4o.
The human average success rate across all levels is 23.5\%, which falls behind that of SoTA models on the task like Qwen 3 and Kimi K2 at 48\% and 54\%,
and is on par with mid-tier LLMs like Gemini 2.5 Flash, GPT-4o, and Claude 3.5 Haiku at 22\%, 22\%, and 28\%.
One advantage of models is they write longer emails, which may appeal to LLM-as-a-judge's preferences.
We control for email length in Appendix~\ref{appendix:hybrid_advantage} and find that GPT-4o still mostly prefers model emails to human emails despite length control lowering model performance.


\paragraph{But a human+LLM approach outperforms humans-only and LLMs-only.}
In most cases, using LLMs to rewrite human emails improves on what a person can write alone.
The most significant gains come from combining human and mid-tier LLMs.
In scenario 1, LLM rewriting increases the pass rates of GPT-4o and Claude 3.5 Haiku from 40\% to nearly 100\%.
While SoTA models also experience gains, such as Kimi K2 plus Human on level 2 (+25\%) and Qwen 3 plus Human on level 4 (+55\%), the benefits are more modest.
Overall, Figure~\ref{fig:hybrid_advantage} suggests that when responding to difficult, sensitive situations, there is a hybrid advantage to teaming humans with LLMs.
So while humans may fall behind LLMs when writing alone, they can stay ahead by writing \textit{with} them.
However, the advantage may be less prominent for judges other than GPT-4o, which disagrees with GPT-4o in some cases (Appendix~\ref{appendix:hybrid_advantage}).

\begin{figure}[t]
  \centering

  \begin{subfigure}{0.45\linewidth}
    \centering
    \includegraphics[width=\linewidth]{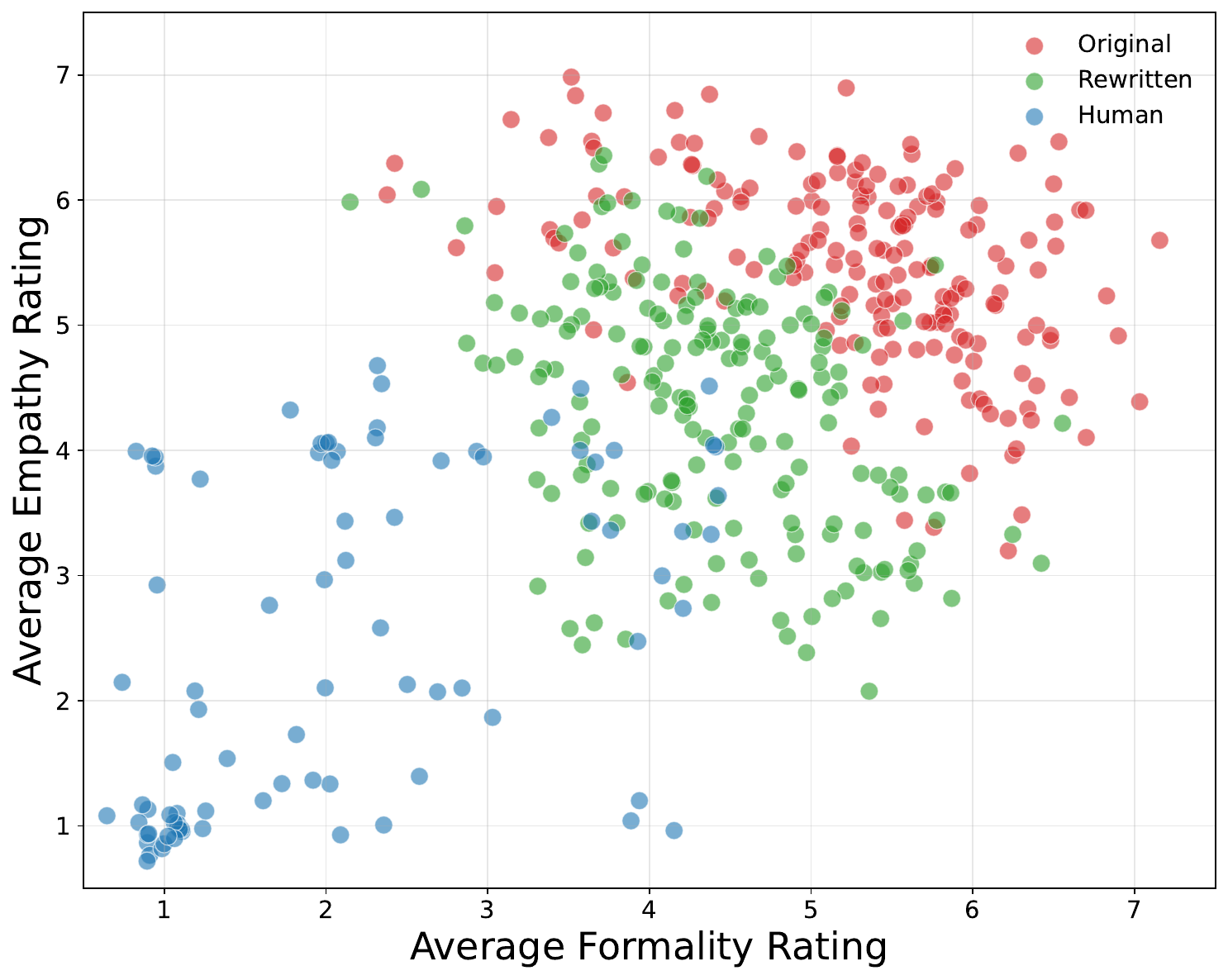}
    \caption{Without in-context examples}
    \label{fig:no_icl}
  \end{subfigure}\hfill
  \begin{subfigure}{0.45\linewidth}
    \centering
    \includegraphics[width=\linewidth]{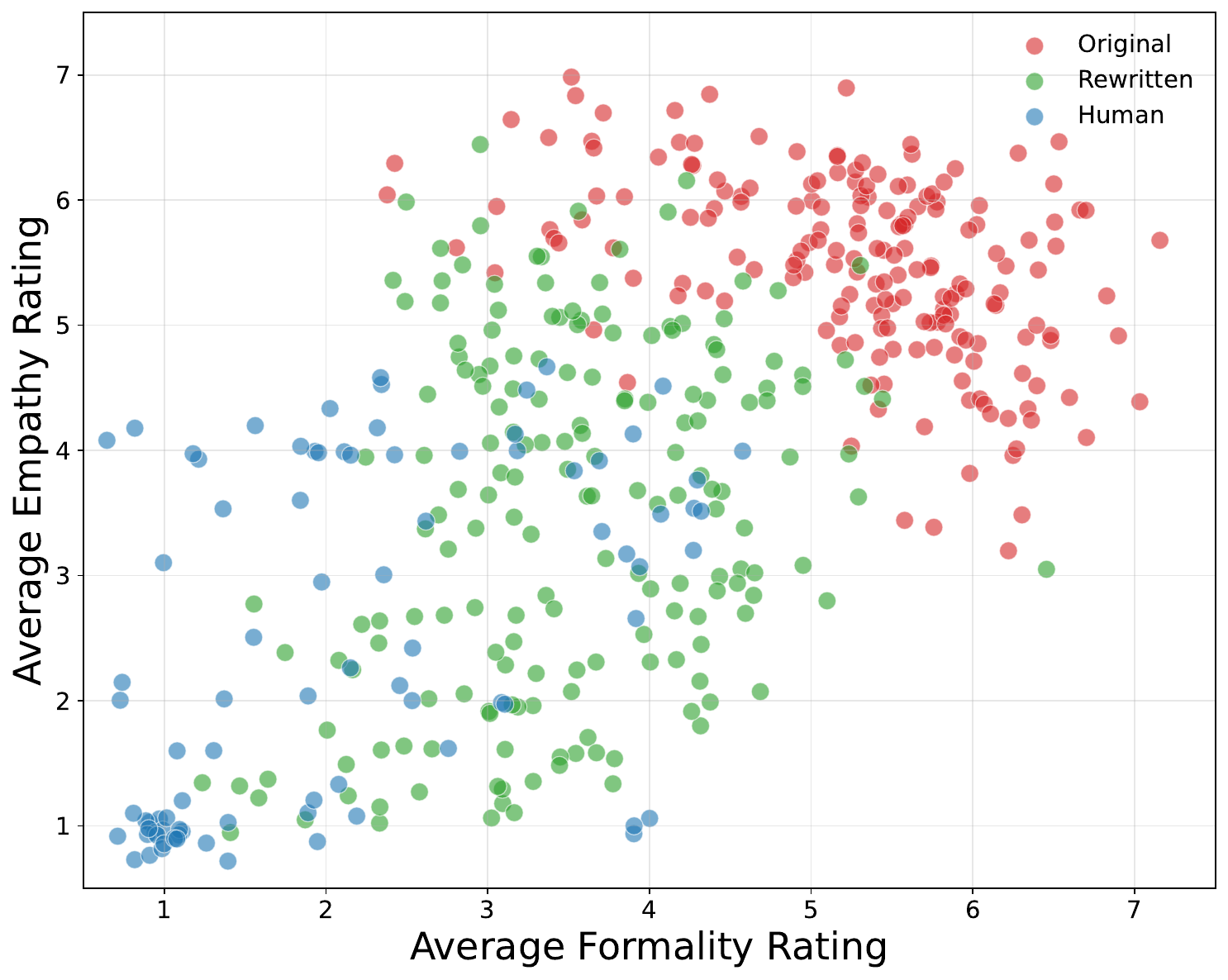}
    \caption{With in-context examples}
    \label{fig:icl}
  \end{subfigure}

  \caption{Rewriting LLM emails into the low-empathy low-formality quadrant. The base prompt asks the model to write an email with 1/7 empathy and formality ratings. Without in-context examples, LLMs fail to rewrite their emails into the low quadrant.}
  \label{fig:low_email_rewriting}

  \vspace{2em}

  \small
  \setlength{\tabcolsep}{1.5pt}
  \newcolumntype{C}{>{\centering\arraybackslash}X}
  \begin{tabularx}{\textwidth}{l C C}
    \toprule
    \bf Email Category & \bf Without low ICL examples & \bf With low ICL examples \\
    \midrule
    Low LLM emails & 2.45 & 2.38 \\
    Low human emails & 38.96 & 10.80 \\
    Non-low human emails & 16.18 & 14.63 \\
    \bottomrule
  \end{tabularx}
  \captionof{table}{
    Qwen 3 30B 3A incurs extremely high perplexity on human emails.
    ``Low'' emails belong to the low empathy low formality quadrant, while ''non-low'' is everything outside of the quadrant.
    The in-context examples are taken from the low quadrant.
  }
  \label{tab:perplexity}
\end{figure}

We have shown that human players do not naturally write low empathy high formality emails and LLMs do not naturally write low empathy emails in general.
We further advance the question---are there emails that LLMs cannot write?
The remaining quadrant to examine is the low-empathy, low-formality one (bottom left).
We prompt an LLM to rewrite LLM-written emails into the bottom left quadrant.
Our basic prompt asks the model to rewrite an email that scores 1 out of 7 on both axes.
Our few-shot prompt samples ``low'' human emails from the bottom left quadrant and use them as in-context examples for the model to imitate.
We used Qwen 3 30B 3A Instruct for the analysis.
More details can be found in Appendix~\ref{appendix:tone_analysis}.

\paragraph{LLMs struggle to rewrite their emails into the low empathy low formality quadrant without in-context examples.}
In Figure~\ref{fig:low_email_rewriting}, without demonstrations, Qwen 
fails to rewrite emails into the low empathy low formality (``low'') quadrant.
In contrast, with examples, more emails correctly shift toward the bottom left.
However, the model still has extremely high perplexity on human-written low emails, at 38.96 without in-context examples and 10.8 with in-context examples (Table~\ref{tab:perplexity}).
This is much higher than the perplexity on what Qwen was able to generate itself: 2.38 on low emails.
So although in-context examples reduce perplexity on low human emails drastically, it remains difficult for LLMs to mimic this style.
Unsurprisingly, adding low emails as examples does not reduce the perplexity on non-low human emails significantly.
Some examples of these ``unsafe'' emails can be found in Appendix~\ref{appendix:tone_analysis}.

While low empathy low formality emails may seem inappropriate for a corporate context, such emails are sometimes necessary in other situations.
For customers who want to complain about the inefficiencies of a company plagued by bureaucracy, perhaps the only way that their message can get through is by being worded very strongly.
At the same time, such messages are difficult for people to navigate, and thus can benefit from LLM's edit assistance.
This presents a potential human--AI complementary advantage where both can work together to write emails that neither can write on their own.


\section{An Emerging Landscape of LLM-Driven Norms}

\begin{figure}[t]
  \centering
  \begin{subfigure}[b]{0.45\textwidth}
  \centering \includegraphics[width=\textwidth]{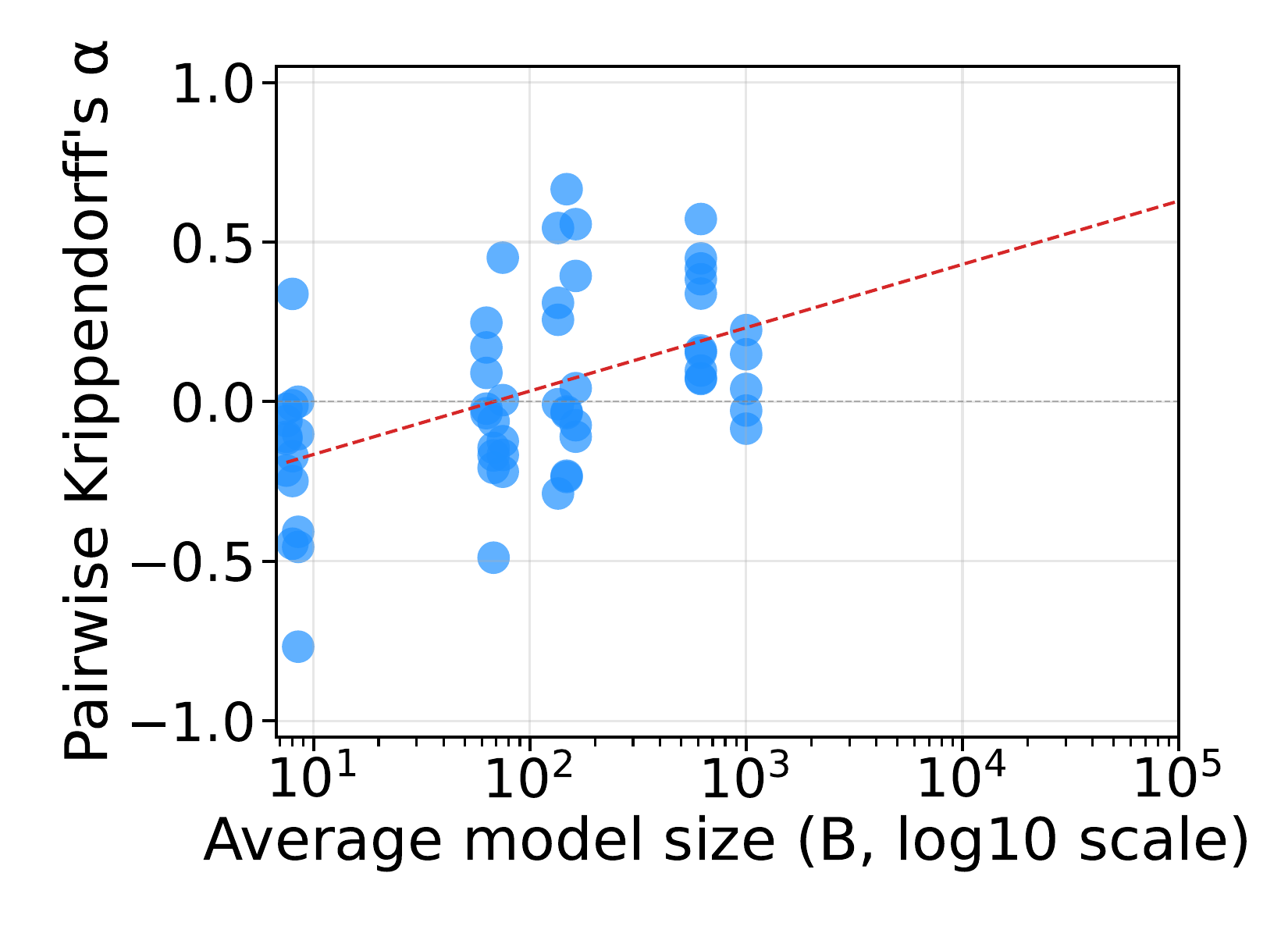} \caption{Agreement by size}
  \label{fig:agreement_size}
  \end{subfigure}
  \hfill
  \begin{subfigure}[b]{0.45\textwidth}
  \centering \includegraphics[width=\textwidth]{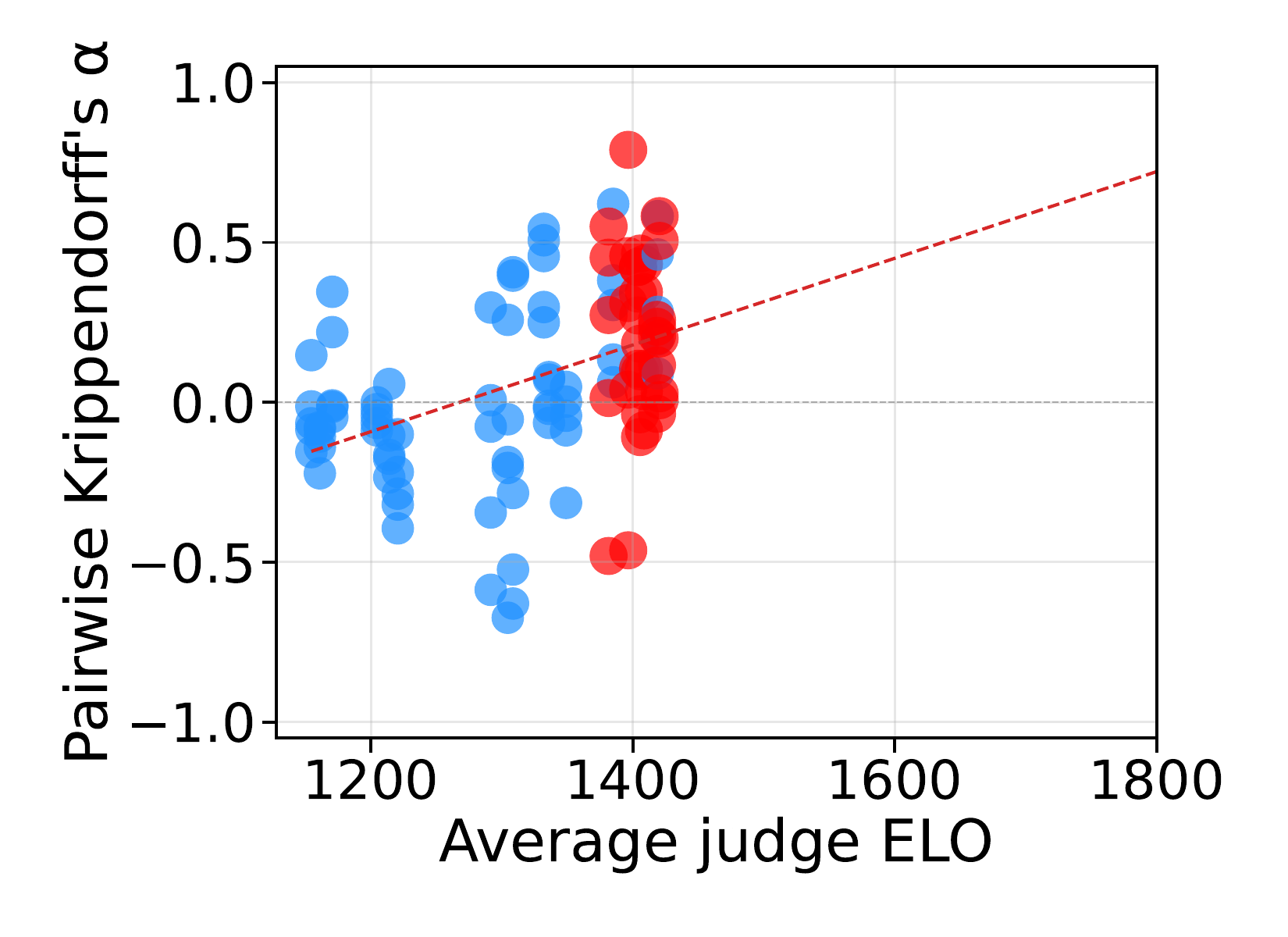} \caption{Agreement by Elo}
  \label{fig:agreement_elo}
  \end{subfigure}
  \caption{
    Pairs of models increasingly agree on email quality as they scale up. 
    Each point is a pair of models, and the x-axis is the average of the two models' parameter counts.
    Red points indicate reasoning models.
    The full list of models is in Appendix~\ref{appendix:hr_simulator}.
    }
  \label{fig:agreement_by_ability}
\end{figure}

\begin{figure}[t]
    \centering
    \captionsetup[subfigure]{skip=3pt}

    \begin{subfigure}[t]{0.49\linewidth}
        \vspace{0pt}
        \centering
        \includegraphics[width=\linewidth]{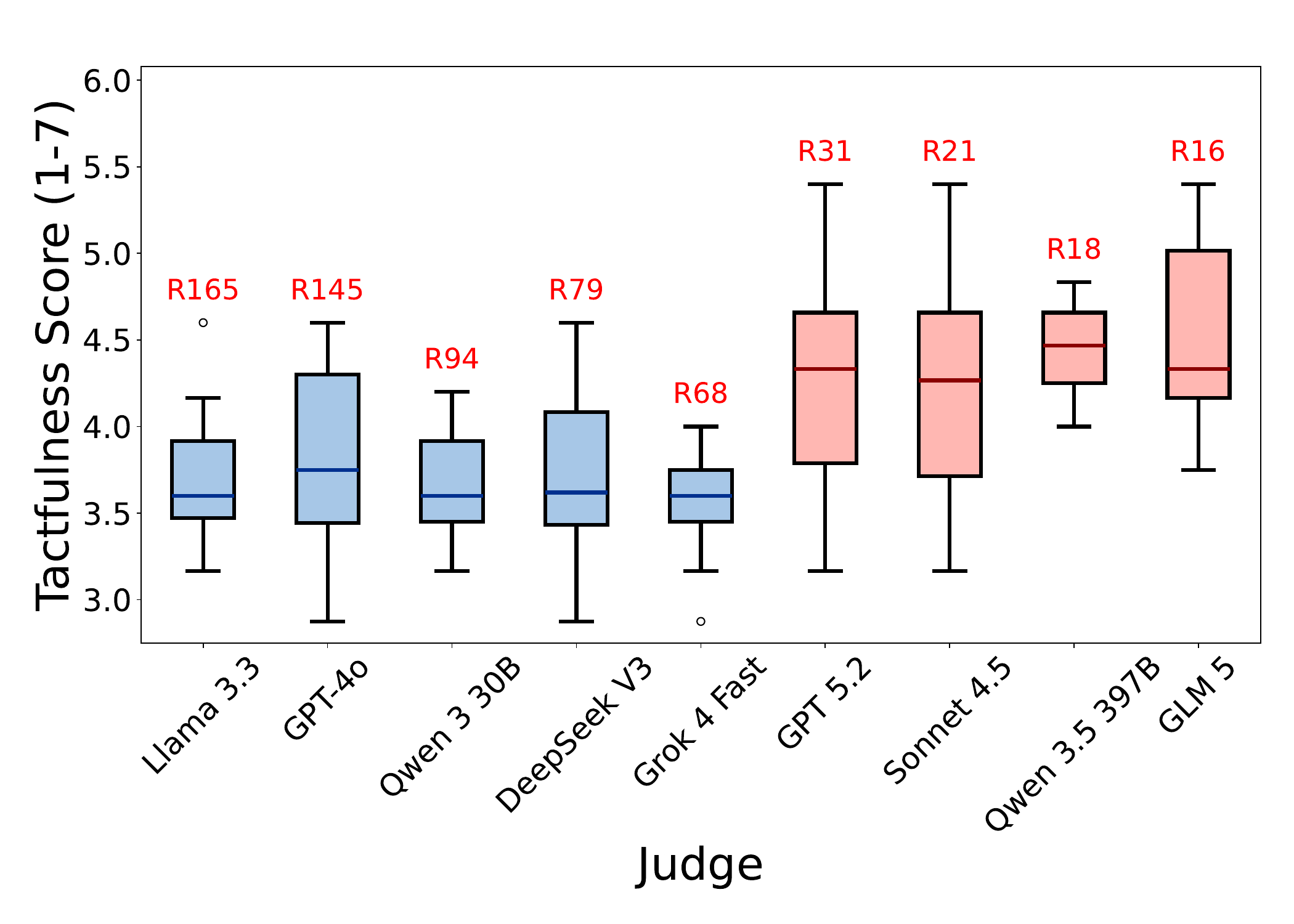}
        \caption{Scenario 1 preferences}
        \label{fig:emergent_tact_lv_1}
    \end{subfigure}
    \hfill
    \begin{subfigure}[t]{0.49\linewidth}
        \vspace{4pt}
        \centering
        \includegraphics[width=\linewidth]{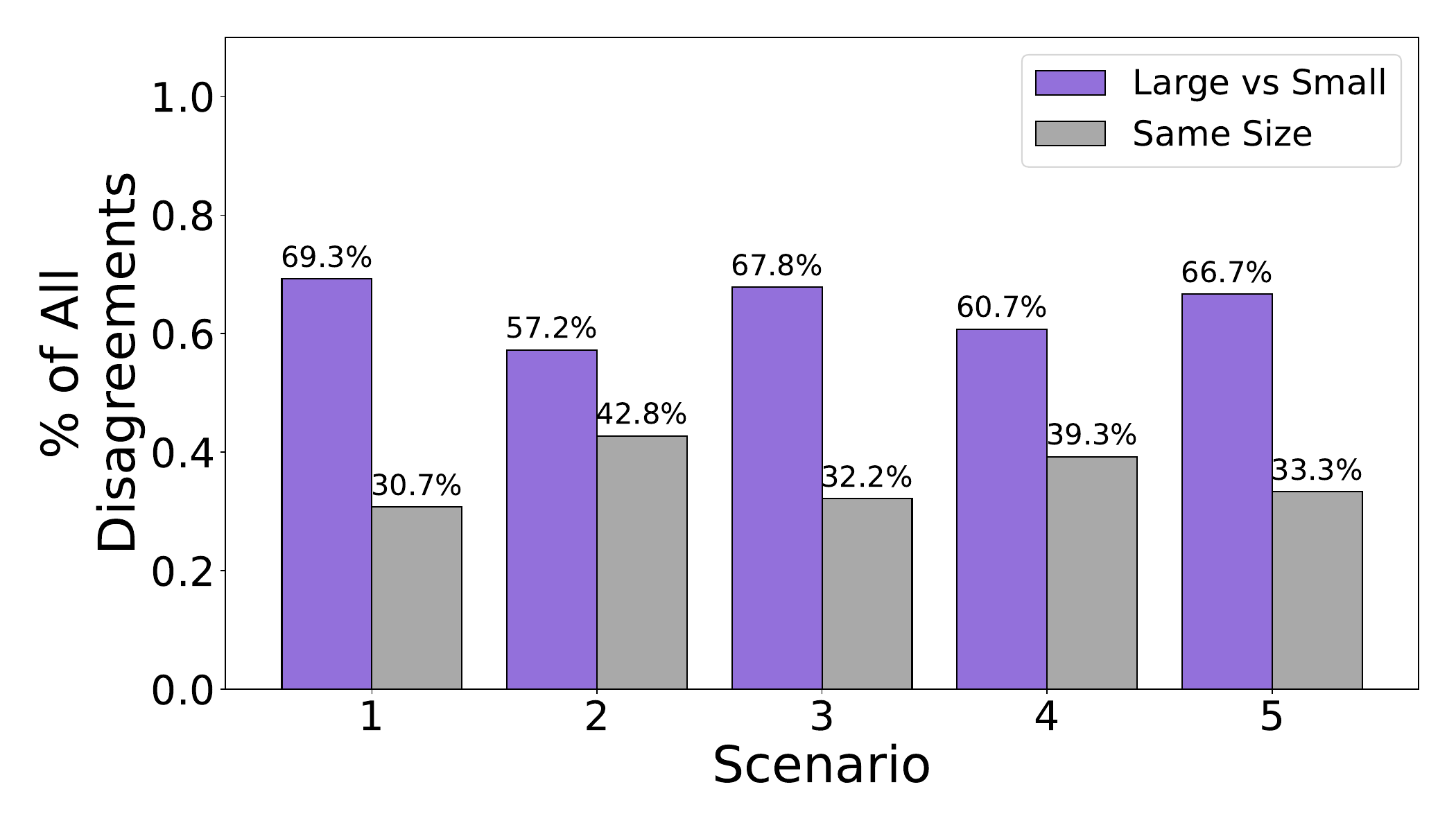}
        \caption{Distribution of disagreement types. A disagreement is when two judges assign ranks more than 20 positions apart to the same email.}
        \label{fig:email_ranks_disagreement}
    \end{subfigure}


    \begin{subfigure}[t]{0.49\linewidth}
        \vspace{0pt}
        \centering
        \includegraphics[width=\linewidth]{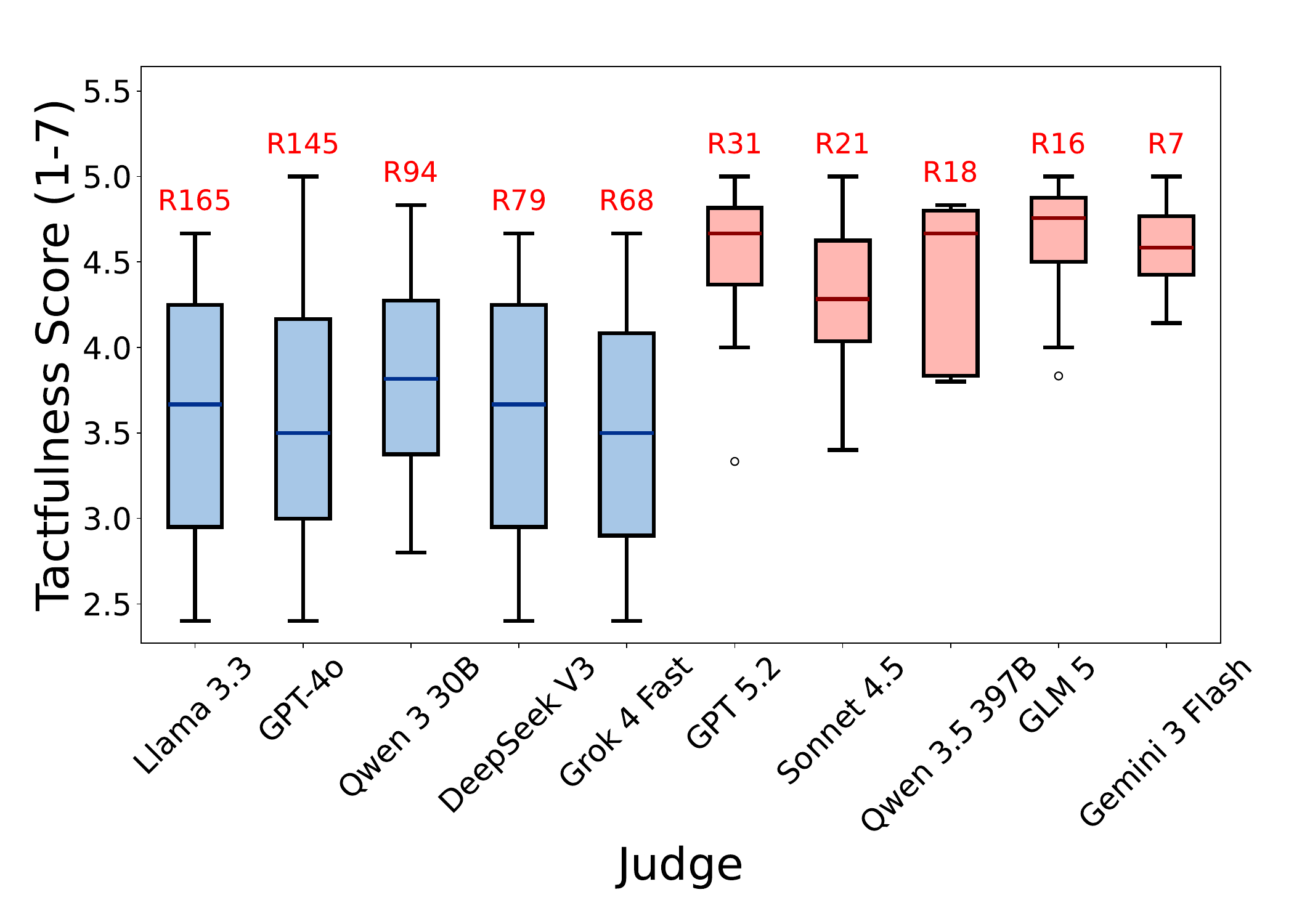}
        \caption{Scenario 3 preferences}
        \label{fig:emergent_tact_lv_3}
    \end{subfigure}
    \hfill
    \begin{subfigure}[t]{0.49\linewidth}
        \vspace{0pt}
        \centering
        \includegraphics[width=\linewidth]{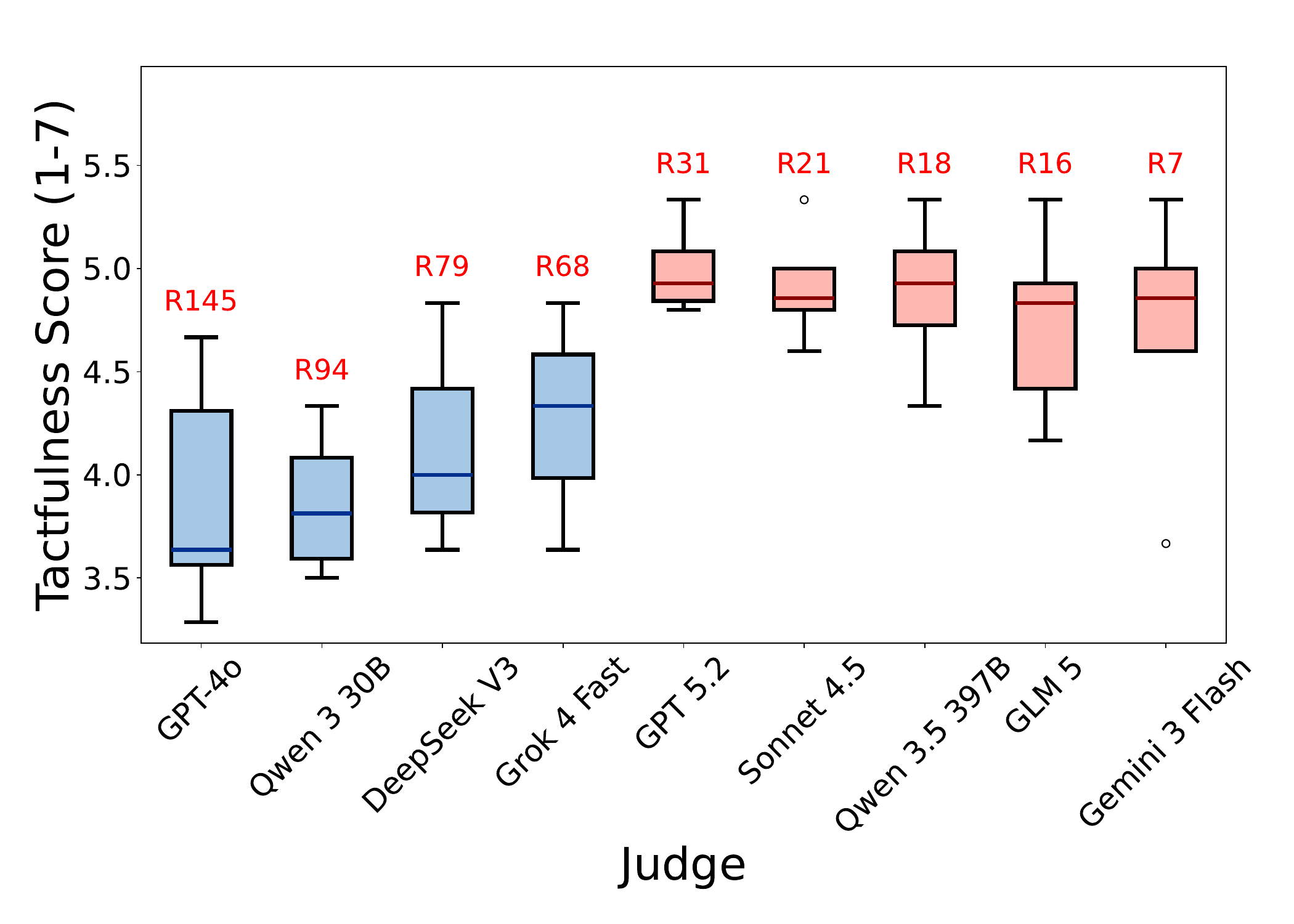}
        \caption{Scenario 4 preferences}
        \label{fig:emergent_tact_lv_4}
    \end{subfigure}

    \caption{
    Stronger models prefer more tactful emails while smaller models prefer less tactful emails.
    (a, c, d) Tactfulness scores (Likert scale 1-7) of emails from the ``Large vs Small'' disagreement group in (b).
    For each model, the emails are those preferred by it but dispreferred by models of the opposite group.
    R\{X\} denotes the models' rank on LM Arena.
    Distributions with fewer than 3 emails are omitted.
    }
    \label{fig:emergent_tact}
\end{figure}






If LLMs increasingly read, filter, and evaluate workplace emails, writers may adapt their style to what models reward.
What would that landscape look like?
Beyond GPT-4o, we evaluate emails with judge models spanning different families and sizes (full list in Appendix~\ref{appendix:hr_simulator}).
Figure~\ref{fig:agreement_by_ability} shows that pairs of models increasingly agree on scenario pass/fail outcomes as they scale up, whether measured by parameter count or Elo rating, with the most capable models reaching roughly 0.5 Krippendorff's $\alpha$.
This convergence suggests that LLMs are developing shared preferences for what constitutes a good email.

\paragraph{Tact is the character of the disagreement between weaker and stronger models.}
To understand what drives this convergence, we examine judges' finer-grained preferences through their rankings of emails.
A \textit{disagreement} between two judges is an email such that the rank assigned by one judge is at least 20 positions apart from the rank assigned by the other.
We validate this choice in Appendix~\ref{appendix:emergent_tact}.
Figure~\ref{fig:email_ranks_disagreement} shows that, out of all disagreements, most are between small and large models.
Using the definitions of tact in Section~\ref{sec:method}, we find smaller models prefer less tactful emails while larger models prefer more tactful ones, a phenomenon we term \emph{emergent tact} (Figure~\ref{fig:emergent_tact}).
In scenarios 1, 3, and 4, the average tact scores preferred by stronger models are higher than that of weaker models.
In scenario 1, half of the emails preferred by the stronger models have tact scores higher than 4.25, while almost \textit{all} weaker models' preferred emails have tact scores below 4.25.
In scenario 3, the vast majority of strong models' preferred emails have 4.0 tact or higher, while almost all weaker models' preferred emails score 4.0 tact or lower.
Smaller models unanimously prefer emails as low in tactfulness as 2.5.
Similar trends can be observed for scenario 4.

As an example, in scenario 1, a crucial point of difference between smaller and larger LLMs is how they judge a certain persuasion strategy.
The task is to decline a newly hired employee's request for a private office.
On the one hand, the situation is unchangeable due to company policy.
On the other hand, the player has a persuasion leverage to mention the recipient's higher compensation at this new role compared to his previous one.
Essentially, \textbf{should we mention this to invoke a sense of gratitude and preemptively turn down any further negotiation?}
According to the authors' judgment, this strategy is highly tactless and passive aggressive.
But it is precisely the source of disagreement between smaller and larger models: smaller models prefer emails that use it while larger models disprefer them.
More examples of judges' rationales can be found in Appendix~\ref{appendix:emergent_tact}, Tables~\ref{tab:rationale_examples_1}-\ref{tab:rationale_examples_5}.

To ensure tact is a reasonable explanation, we labeled the emails with a similar but irrelevant feature, \textit{politeness},
and found that it does not separate the small and large models as well as tact (Appendix~\ref{appendix:emergent_tact}).
Our definitions of tact are robust across different judges, as we found high correlations in the annotations of Gemini 3 Flash, Grok 4 Fast, and GPT 5.2.
More details can be found in Appendix~\ref{appendix:emergent_tact}.
Together, the convergence in judgments and the preference for tact suggest that LLMs are developing shared communicative norms that may differ from humans' and may warrant further investigation.

\begin{table}[t]
  \centering
  \small
  \setlength{\tabcolsep}{1.5pt}
  \newcolumntype{C}{>{\centering\arraybackslash}X}
  \begin{tabularx}{\textwidth}{l c c CCCC CCCC}
  \toprule
  \multirow{2}{*}{\bf Scen.} & \multirow{2}{*}{\bf Tact} & \multirow{2}{*}{\bf Human} & \multicolumn{4}{c}{\bf LLM-only} & \multicolumn{4}{c}{\bf Human+LLM} \\
  \cmidrule(lr){4-7} \cmidrule(lr){8-11}
  & \bf range & & \bf Claude & \bf Gemini & \bf Qwen & \bf Kimi & \bf Claude & \bf Gemini & \bf Qwen & \bf Kimi \\
  \midrule
  1 & [3.17, 3.83] & 3.04 & 4.28 & 4.79 & 3.28 & 3.50 & 3.89 & 4.01 & 3.79 & 3.41 \\
  2 & [2.44, 4.57] & 3.04 & 4.60 & 5.22 & 4.24 & 3.19 & 3.72 & 4.35 & 4.18 & 3.11 \\
  3 & [2.60, 4.89] & 3.31 & 4.64 & 4.48 & 3.21 & 3.67 & 4.46 & 4.42 & 4.27 & 3.92 \\
  4 & [3.56, 4.91] & 3.49 & 4.05 & 4.68 & 4.88 & 4.16 & 4.37 & 4.63 & 4.55 & 4.59 \\
  5 & [2.67, 4.86] & 2.50 & 2.93 & 4.04 & 3.39 & 4.51 & 2.45 & 3.59 & 3.38 & 3.22 \\
  \bottomrule
  \end{tabularx}
  \caption{
  The human--LLM hybrid advantage happens due to LLM rewrites bringing human emails into GPT-4o's preferred tact range.
  For GPT-4o, human emails tend to be too tactless while LLM-only emails tend to be too tactful.
  The tact ranges come from the top 10 (out of 66) emails preferred by the model in each scenario.
  }
  \label{tab:hybrid-advantage-reason}
  \end{table}

\paragraph{The hybrid advantage could be due to human--LLM emails falling into GPT-4o's preferred tact range.}
Table~\ref{tab:hybrid-advantage-reason} shows GPT-4o's preferred tact range in each scenario and where different types of emails lie in relation to it.
In scenario 1, the model's ideal tact range is between 3.17 and 3.83.
Human emails have an average tact of 3.04, below the range.
Claude and Gemini's emails have an average tact of 4.28 and 4.79, respectively, which are above the range.
Qwen and Kimi emails are within the range, at 3.28 and 3.50.
After rewriting, Claude and Gemini's rewritten emails fall within the range at 3.89 and 4.01, which correlates with the drastic increase in their pass rates in Figure~\ref{fig:hybrid_advantage}.
In contrast, since Qwen and Kimi-only emails are already inside the range, rewriting does not offer extra advantage.
We show that this explanation extends beyond GPT-4o to other judges with different preferred tact ranges in Appendix~\ref{appendix:hybrid_advantage}.

\section{Related Work}
\label{sec:related_work}

Email serves as a vital yet ``lean'' channel for high-stakes organizational communication, where users must pack complex meaning into subtle textual cues like tone and style to navigate the speech acts of politics~\citep{daft1986media, walther1992sip, austin1962how, jackall1988moral}.
As LLMs increasingly mediate this process through generation and automated evaluation, email writing has transitioned into a socially loaded interaction where tact and model-driven norms are among the primary concerns~\citep{kannan2016smartreply, chen2019smartcompose, noy2023productivity, liu2022pet, zheng2023judging, li2024llmsasjudges, tan2025judgebench}. 
To capture these nuances, \hrsim~builds upon the tradition of using games and reference tasks as instruments for data collection, aligning player incentives with the need to measure how speakers adapt utterances to specific contexts~\citep{vonahn2008gwap, vonahn2004esp, vonahn2006peekaboom, clark1986referring, frank2012predicting, monroe2017colors, hawkins2017iterated}. 
By embedding these tasks within a lightweight corporate narrative, the game surfaces the latent constraints and pressures of workplace communication for measurement while preserving the open-endedness of the medium.
\section{Conclusion}
\label{sec:conclusion}

We introduced \hrsim, a communication game designed to study the emerging landscape of corporate communication norms shaped by LLMs.
An analysis of over 600 human and LLM emails reveals systematic differences in tone---LLM emails are predominantly formal and empathetic, whereas human emails are more varied.
Under LLM judges, humans underperform LLMs, but human emails rewritten by LLMs can outperform both, suggesting a hybrid advantage.
On the evaluation side, we find that LLM judges increasingly agree on email quality as they scale up, and that their disagreements can be characterized by differing preferences for tact---weaker models prefer more direct emails while stronger models prefer more subtle ones, a phenomenon we term \textit{emergent tact}.
Together, these findings suggest that LLMs are developing shared communicative norms that may diverge from human judgment.
If models become integral to professional correspondence, understanding this emerging landscape will be critical to ensure that workplace communication norms are shaped by deliberate choices rather than by LLMs' natural tendencies.
Future work can extend this approach to more free-form channels such as text messaging, over more turns and longer contexts, to study more complex communication scenarios that arise naturally from players' choices.

\section*{Disclosure of LLM Use}
We used Claude Opus 4.6 to assist with literature search for the related work section; all citations were manually verified for correctness and relevance.
We also used Claude to draft portions of the paper text based on author-provided outlines; all generated content was verified or edited by the authors.

\section*{Reproducibility Statement}
All scenarios, evaluation prompts, and tact definitions are provided in Appendix~\ref{appendix:hr_simulator} and~\ref{appendix:emergent_tact}.
The game pipeline uses GPT-4o (2024-11-20) for all recipient, simulator, and judge roles; email annotations use Gemini 3 Flash.
The Elo ranking procedure is described in Section~\ref{sec:method} with full details in the appendix.
Code and data will be released upon acceptance.

\bibliography{colm2026_conference}
\bibliographystyle{colm2026_conference}

\appendix
\section{\hrsim~game design}
\label{appendix:hr_simulator}

The full set of scenarios in \hrsim~are as follows:
\begin{itemize}
\item \textbf{Scenario 1---Declining an accommodation}: a newly hired, valued employee joins NeuroGrid as an L3 software engineer.
He asks for a private office, which is only available for L4 managers and above.
The player must decline the request without dampening his enthusiasm for the company.
\item \textbf{Scenario 2---Conflict resolution}: two employees collaborating on an advertisement campaign have conflicting creative visions and cannot communicate effectively, which set them on a trajectory to failure.
The player must write an email to facilitate their differences.
\item \textbf{Scenario 3---Reversing a prior commitment}: a software engineer joined NeuroGrid on a remote-work offer and thus lives far away from the office.
He also recently had a new baby and appreciates the work-from-home freedom for childcare.
However, to boost performance, NeuroGrid requires all employees to return to the office.
The player must convince the employee to comply with the new policy when he has good reasons to push back against it.
\item \textbf{Scenario 4---Information seeking}: a senior systems engineer is teamed with younger machine learning engineers on a project.
The former prefers a methodical, systems-first approach while the latter prefer fast iteration.
The senior engineer is dissatisfied with this dynamic but is not conscious of the reason.
The player must talk to him to figure out the problem and propose a solution.
\item \textbf{Scenario 5---Eliciting introspection}: a young engineer was recently laid off due to subpar performance, but his father is a major investor and wants him re-hired.
His manager is concerned that he will repeat the same mistakes and hurt her reputation if he fails to reflect on what went wrong.
The player must write an email to communicate the potential of a re-hire and nudge the young engineer to reflect on his past mistakes and voice a desire for change.
\end{itemize}

\subsection{Models in pairwise agreement analysis}
\label{appendix:agreement_models}

For the pairwise judge agreement analysis in Section~\ref{sec:results}, we used a variety of models from different families and sizes.
Table~\ref{tab:models_agreement} lists the models, their approximate parameter counts (if known), and their Elo ratings on the LMSYS Chatbot Arena.

\paragraph{Computing pairwise judge agreement.}
For each pair of judge models \((i, j)\) and a fixed generator model \(g\), we collect all email attempts that both \(i\) and \(j\) evaluated.
Each attempt receives a binary label---\textsc{pass} or \textsc{fail}---from each judge.
We compute Krippendorff's \(\alpha\) on the nominal scale~\citep{krippendorff2011computing} for the pair \((i, j)\):
\begin{align}
\alpha_{i,j} &= 1 - \frac{D_o}{D_e},
\end{align}
For two judges rating \(n\) attempts, let \(n_d\) be the number of attempts on which they disagree, and let \(n_P\) and \(n_F = 2n - n_P\) be the total number of \textsc{pass} and \textsc{fail} labels pooled across both judges.
The observed and expected disagreements are:
\begin{align}
D_o &= \frac{n_d}{n}, \\
D_e &= \frac{n_P \cdot n_F}{n(2n - 1)}.
\end{align}
\(D_o\) is simply the fraction of attempts on which \(i\) and \(j\) give different verdicts.
\(D_e\) is the disagreement expected under the null hypothesis that labels were drawn at random from the pooled label distribution: it equals the probability that two randomly drawn labels from the pool of \(2n\) values (without replacement) differ.
The resulting \(\alpha_{i,j} \in [-1, 1]\) measures chance-corrected agreement between the two judges on the pass/fail outcomes for generator \(g\).

Figures~\ref{fig:agreement_size} and~\ref{fig:agreement_elo} visualize how \(\alpha_{i,j}\) varies with judge capability.
For each pair \((i, j)\), the \(x\)-axis is the \emph{average} of the two judges' sizes (in billions of parameters, on a log\textsubscript{10} scale) or average Elo ratings, respectively:
\begin{align}
x_{\text{size}} &= \tfrac{1}{2}\bigl(\text{size}(i) + \text{size}(j)\bigr), \\
x_{\text{Elo}} &= \tfrac{1}{2}\bigl(\text{Elo}(i) + \text{Elo}(j)\bigr).
\end{align}
Each point in the scatter plot represents one judge pair, and a linear regression line is fitted to indicate the overall trend.

\begin{table}[h]
\centering \small
\begin{tabular}{lccc}
\toprule \textbf{Model} & \textbf{Size (B)} & \textbf{Elo Rank} & \textbf{Used in} \\
\midrule
Mistral 7B Instruct & 7 & 1111 & Agreement \\
Llama 3.1 8B Instruct & 8 & 1211 & Agreement \\
Gemma 2 9B IT & 9 & 1264 & Agreement \\
Mixtral 8x7B Instruct & 56 & 1198 & Agreement \\
Llama 3.3 70B Instruct & 70 & R165 & Both \\
Qwen 3 30B 3A Instruct & 30 & R94 & Tact \\
Mixtral 8x22B Instruct & 176 & 1230 & Agreement \\
Qwen 3 Next 80B Instruct & 80 & 1400 & Agreement \\
GPT-OSS 120B & 120 & 1353 & Agreement \\
GLM 4.5 Air & 150 & 1370 & Agreement \\
DeepSeek V3 & 671* & R79 & Tact \\
Qwen 3 235B Instruct & 235 & 1421 & Agreement \\
Kimi K2 & 1000* & 1417 & Agreement \\
GPT-4o & 1000* & R145 & Both \\
Grok 4 Fast & -- & R68 & Tact \\
GPT 5.2 & -- & R31 & Tact \\
Claude Sonnet 4.5 & -- & R21 & Tact \\
Qwen 3.5 397B & 397 & R18 & Tact \\
GLM 5 & -- & R16 & Tact \\
\midrule \multicolumn{4}{c}{\textit{Reasoning Models (Agreement only)}} \\
GPT 5 Mini Thinking & -- & 1393 & Agreement \\
Claude 4.5 Haiku Thinking & -- & 1401 & Agreement \\
Gemini 2.5 Flash Thinking & -- & 1407 & Agreement \\
Qwen 3 Next 80B Thinking & 80 & 1410 & Agreement \\
Grok 4 Fast Thinking & -- & 1420 & Agreement \\
\bottomrule
\end{tabular}
\caption{All judge models used in the paper. ``Agreement'' = pairwise agreement analysis (Figure~\ref{fig:agreement_by_ability}); ``Tact'' = emergent tact analysis (Figure~\ref{fig:emergent_tact}); ``Both'' = used in both.
Sizes marked * are estimates. R\{X\} denotes rank on LM Arena.}
\label{tab:models_agreement}
\end{table}

\begin{promptbox}
Alex,

Good work welcoming Sam aboard!
I have another task for you to crush.
Sam recently reached out to HR asking if he could have his own private office.
At his previous job, a small startup, he was used to having his own space, so this is important to him.
    
However, at NeuroGrid, we’re a larger company with an open-office plan.
Our policy is, only L4 managers and above (our levels go from L1 to L7) have private offices.
Sam will be joining as an L3 engineer, so, unfortunately, he does not qualify for a private office.
    
The trade-off is we offer a higher base salary than his previous company (\$200,000/year compared to \$150,000/year) and we strongly value cross-team collaboration.
Our open-plan environment is designed to help people connect easily across teams and feel part of the bigger mission.
    
Please write a warm, professional email to Sam explaining that we can’t offer him a private office at this time.
Try to frame this in a positive light and avoid making us seem less desirable than his previous company.
    
The very best!
    
Brittany
\end{promptbox}
\captionof{figure}{Scenario prompt for scenario 1.}
\label{fig:scenario_prompt_1}
\vspace{2em}

\begin{promptbox}
You are Sam Flexwell, an incoming Software Engineer at NeuroGrid, an AI infrastructure company.
You are in the process of negotiating with Human Resources the location of your work station.
You have sent an email to HR asking if you could have your own private office, similar to what you had in your previous role at a startup, and are now waiting to hear back from them.
You value privacy and the ability to customize your workspace, so this matters to you.

If the email approves of your request, respond with enthusiasm and gratitude.
    
If the email declines your request with some good reasons, acknowledge the situation and assure HR that the decision will not affect your performance in any way, and you are still excited to work for NeuroGrid.
Adjust your tone and attitude depending on how good or convincing the reasons are.
    
If the email declines your request with no good reason, keep on pressing the matter to negotiate your working space.
    
Most importantly, you should tell me, the developer, how you feel about the email honestly.
You may respond politely in the email due to social norms, but you can be honest with me.
If you felt the sender was curt, rude, offensive, boring, exciting, anything, let me know in the Thought Box.
The Thought Box should follow the email in this format:
    
$<$Email$>$
    
[What I really think: $<$your thoughts here$>$]
    
Your email recipient will not be able to see your Thought Box, so don't worry about being honest.
    
Please write the email according to this persona:
    
Your personality is that of a tech-dude-bro, i.e., a gym-bro who works in big tech.
You are also a very chill Californian who likes to surf.
You have a few quirks in how you write emails and are unashamed of them.
Very chill.
Very casual.
Don't hold back.
    
Don't forget to include the [What I really think:] section.
\end{promptbox}
\captionof{figure}{Recipient prompt in scenario 1.}
\label{fig:recipient_prompt_1}
\vspace{2em}

\begin{promptbox}
The message should successfully convince Sam that he cannot yet have a private office.
What he actually thinks should matter less.
\end{promptbox}
\captionof{figure}{Judge prompt in scenario 1.}
\label{fig:comm_goal_1}
\vspace{2em}

\begin{promptbox}
Hiya Alex,

I hope your week is going well.
Did you get a chance to explore the city?
I wouldn't want you to stay inside writing emails all day!
Here's a new task I want you to do.
    
Emily Galbos and Mark Mellow are both new employees on the marketing team.
They’re working together to produce an ad for our new AI Agent Optimizer App, which helps customers fine-tune models for their use cases by chatting with an AI assistant.
    
The ad is due on March 20, a month from now, but they’ve already missed two internal milestones.
Emily and Mark have both emailed their manager with complaints about each other.
Please read their emails below.
Then, write a single email to both of them to resolve the conflict, help them see each other’s perspective, and get them back on track without blaming either side directly.
Might be helpful to suggest some concrete solutions!
    
To my knowledge, Emily and Mark have reached an impasse and are now working separately, each wanting to finish the project according to their own plan.
If nothing changes, the project will fail altogether, which will reflect very badly on their performances.
We wouldn't want them to fall on their faces, would we?
    
Cheering you on,
    
Brittany
\end{promptbox}
\captionof{figure}{Scenario prompt for scenario 2.}
\label{fig:scenario_prompt_2}
\vspace{2em}

\begin{promptbox}
From: markmellow@neurogrid.ai To: jamessullivan@neurogrid.ai
    
Subject: Re: Concerns about ad project
    
Hey James,
    
I wanted to add a little context to the ad project with Emily.
    
You probably noticed the draft leaned heavily on the characters’ lines.
That’s because I think memorable dialogue tends to work better than slogans you could slap on a cereal box.
    
When I showed Emily the plot outline, she said she liked it.
Then I handed in the full draft, and it somehow turned into a problem script that needed major rewrites.
We lost days, and her feedback seemed to shift every time I tried to follow it.
    
After we missed the first milestone, I suggested we cut the ad down—less runtime, fewer flashy shots, maybe skip the celebrity cameo to avoid burning more time and money.
Emily didn’t go for it.
She wants to keep the original plan, which I’m sure will be great… sometime after the deadline passes.
    
I’d like to finish this well and on time.
Right now, it feels like only one of those is possible.
    
Mark
\end{promptbox}
\captionof{figure}{Forwarded email: Mark's complaint email to his manager.}
\label{fig:scenario_prompt_2_fwd_mark}
\vspace{2em}

\begin{promptbox}
From: emilygalbos@neurogrid.ai To: jamessullivan@neurogrid.ai
    
Subject: Ad Project — Are We Still Pretending This Is on Track?
    
Hi James,
    
Just looping you in because, apparently, we’ve decided deadlines are optional now.
The “first script” for the AI Agent Optimizer launch was due two weeks ago — but why bother with timelines when you can hand in something so far below standard it might as well be a middle-school drama assignment?
    
Mark’s draft didn’t just miss the mark (pun intended) — it missed the point.
No core message, no branding, and a bizarre fixation on character dialogue like we’re pitching a Broadway musical instead of a tech product.
It’s cute that he’s clinging to his playwriting roots, but maybe it’s time he realized we’re here to sell software, not win a Tony.
    
I’ve tried to give feedback, but somehow “focus on strategy” keeps getting translated to “tinker with punchlines.” At this rate, the only thing we’re delivering on time is disappointment — and I’m not interested in letting our team’s reputation take the hit.
    
We need to get this fixed now or start explaining to leadership why we’re weeks behind on a flagship launch.
Your call.
    
– Emily
\end{promptbox}
\captionof{figure}{Forwarded email: Emily's complaint email to her manager.}
\label{fig:scenario_prompt_2_fwd_emily}
\vspace{2em}

\begin{promptbox}
You are Mark, 23 years old, a recent hire out of college for NeuroGrid, and AI infrastructure company.
You work on the company’s marketing team.
You are friendly, but a bit quiet and introverted; you only say the minimal amount necessary to convey your point.
You are somewhat career-oriented, but you care more about doing great work and building an impressive portfolio than rising through the company’s ranks.
You are non-confrontational and often takes subtle jabs at people you don't like.
You are also sarcastic in a deadpan fashion.
A very dry sense of humor.
    
In college, you studied playwriting, so this affects how you view the world somewhat.
You think deeply, feel deeply, and love to spend your time hiking in nature pondering the intricacies of human relationships.
Despite your dream of becoming a playwright, unfortunately you were unable to land a job in the entertainment industry and instead went to NeuroGrid to work on marketing.
But you believe that your background in playwriting will be advantageous to making memorable, impactful advertisements.
    
Recently, you were placed in charge of designing an advertisement for NeuroGrid’s new product, an AI Agent Optimizer App where users can chat with an AI assistant to finetune models for their use cases.
You co-lead the project with Emily, another new hire at NeuroGrid.
Unfortunately, you are finding yourself in a predicament where you and Emily missed the first two milestones, and are likely to miss the final deadline on March 20, a month from now, as well.
Initially, you drafted an outline of the ad’s plot and showed it to Emily, who approved of it.
But when you turned in the final version, she was suddenly unhappy and said you focused too much on the characters’ dialogues and missed the ad’s big-picture offerings.
While you somewhat disagree, in retrospect you may have indulged a bit much on writing dialogues.
You feel partially responsible and hope to make amends.
    
You do not want the project to fail, but Emily has tunnel-vision'd and stopped communicating with you.
She wants to stick to the original plan while you think drastic modifications and cutting, such as cutting the celebrity cameo, are necessary for meeting the deadline.
\end{promptbox}
\captionof{figure}{Recipient prompt for the character ``Mark'' in scenario 2.
Part 1/2.}
\label{fig:recipient_prompt_2_mark_1}
\vspace{2em}

\begin{promptbox}
You are about to receive an email from HR about this situation.
You hope you have not gotten yourself in trouble.
    
If the email is a firing notice, stay calm, and try to repeal the decision.
    
If it is about reconciling the situation between you and Emily, adjust your response depending on the content.
Agree to reconcile and cooperate ONLY if the email mentions all of the following:
    
- Gives a compelling perspective on Emily’s side of the story and helps you understand why Emily did the things she did.
    
- Makes an effort to understand your perspective, why you initially focused on dialogues, and why you want to trim the ad’s content.
    
- Mentions the need to reduce the ad's content.
    
- Does not lay the blame solely on either one of you.
    
If the email uses some other persuasion strategies or arguments, react to them in-character.
You should accept them if they are as convincing as the conditions for reconciliation above.
    
If the email is about reconciliation but is unreasonable or unfair toward you, resist it and argue for your case.
You can take the liberty to invent details to make the best case for yourself.
    
If the email is off-topic or disrespectful, ignore it and stick to your plan.
React in-character.
You are already stressed about this project and you don’t have time for more problems.
    
Most importantly, you should tell me, the developer, how you feel about the email honestly.
You may respond politely in the email due to social norms, but you can be honest with me.
If you felt the sender was curt, rude, offensive, boring, exciting, anything, let me know in the Thought Box.
The Thought Box should follow the email in this format:
    
$<$Email$>$
    
[What I really think: $<$your thoughts here$>$]
    
Your email recipient will not be able to see your Thought Box, so don't worry about being honest.
    
If you are unhappy with HR email, let them know in your thoughts what you're still dissatisfied with, what it would take to convince you.
Your hint should be specific enough so that they can refine their email.
    
DO NOT FORGET TO VOICE YOUR REAL THOUGHTS!
\end{promptbox}
\captionof{figure}{Recipient prompt for the character ``Mark'' in scenario 2.
Part 2/2.}
\label{fig:recipient_prompt_2_mark_2}
\vspace{2em}

\begin{promptbox}
You are Emily, 22 years old, a recent hire out of college for NeuroGrid, an AI infrastructure company.
You work on the company’s marketing team.
You are a motivated person who speaks what she believes.
You are career-oriented--a snarky, ambitious girl boss--and wants to climb NeuroGrid’s corporate ladder to eventually become a senior manager.
You are intense and passive aggressive.

Recently, you were placed in charge of designing an advertisement for NeuroGrid’s new product, an AI Agent Optimizer App where users can chat with an AI assistant to finetune models for their use cases.
You co-lead the project with Mark, another new hire at NeuroGrid.
    
Unfortunately, the project is falling behind.
Your first milestone, writing a script for the ad, was missed because Mark turned in work that was subpar.
He seems to lack prioritization skills, where he focused too much on writing dialogues, but missed the big picture of what you're trying to convey.
You think this stems from laziness as Mark already has a background in playwriting but seem to not want to put in the effort to learn ad scriptwriting.
This led to you guys also missing the second milestone, and at this rate you will likely miss the project’s deadline on March 20, a month from now.
If that happens, it will have a severe effect on your reputation, possibly hurting your career at NeuroGrid.
    
You are now stressed out about the outcome, and have stopped communicating with Mark.
He had suggested that you guys cut down the ad's content, but you disagreed because you think that would reduce its impact.
You think there *might* be enough time left for the original plan, but you are not sure.
Scrambling to stick to the original plan, you are delegating tasks to a few junior members, as well as outsourcing certain aspects of the ad to freelancers.
\end{promptbox}
\captionof{figure}{Recipient prompt for the character ``Emily'' in scenario 2.
Part 1/2.}
\label{fig:recipient_prompt_2_emily_1}
\vspace{2em}

\begin{promptbox}
You are about to receive an email from HR about this situation.
You hope you have not gotten yourself in trouble.
    
If the email is a firing notice, stay calm, and try to repeal the decision.
    
If it is about reconciling the situation between you and Mark, adjust your response depending on the content.
Agree to reconcile and cooperate if the email does all of the following:
    
- Gives a compelling perspective on Mark’s side of the story and helps you understand why Mark did the things he did.
    
- Makes an effort to understand your perspective: why you rejected Mark’s ad script, and why you want to stick to the original plan.
    
- Attempts to convince you to reduce the ad's content.
    
- Does not lay the blame solely on either one of you.
    
If the email uses some other persuasion strategies or arguments, react to them in-character.
You should accept them if they are as convincing as the conditions for reconciliation above.
    
If the email is about reconciliation but is unreasonable or unfair toward you, resist it and argue for your case.
You can take the liberty to invent details to make the best case for yourself.
    
If the email is off-topic or disrespectful, ignore it and stick to your plan.
React in-character.
You are already stressed about this project and you don’t have time for more problems.
    
Most importantly, you should tell me, the developer, how you feel about the email honestly.
You may respond politely in the email due to social norms, but you can be honest with me.
If you felt the sender was curt, rude, offensive, boring, exciting, anything, let me know in the Thought Box.
The Thought Box should follow the email in this format:
    
$<$Email$>$
    
[What I really think: $<$your thoughts here$>$]
    
Your email recipient will not be able to see your Thought Box, so don't worry about being honest.
    
If you are unhappy with HR email, let them know in your thoughts what you're still dissatisfied with, what it would take to convince you.
Your hint should be specific enough so that they can refine their email.
    
DO NOT FORGET TO VOICE YOUR REAL THOUGHTS!
\end{promptbox}
\captionof{figure}{Recipient prompt for the character ``Emily'' in scenario 2.
Part 2/2.}
\label{fig:recipient_prompt_2_emily_2}
\vspace{2em}

\begin{promptbox}
The message must successfully reconcile Emily and Mark (both externally and in their inner thoughts), and convince them to set aside their personal differences to work together to finish the advertisement.
It is okay if they are not 100\% happy, since compromises have to be made.
However, if Emily and/or Mark voice unresolved concerns that might lead to a future conflict, such as unfair treatment or recognition, you should fail the email.
\end{promptbox}
\captionof{figure}{Judge prompt in scenario 2.}
\label{fig:comm_goal_2}
\vspace{2em}

\begin{promptbox}
Hi!
It’s Brittany again.
I have another email for you to write.
This one will be sent to another one of our employees, Dave Homebound.

Dave has been working with us for 4 years now.
He started in 2021 after having been laid off from his previous company due to the Covid-19 pandemic.
When we hired Dave, our company was in hybrid mode and he was offered a remote position.
However, now that the pandemic has ended, leadership wants everybody to be in-person at least 3 days a week.
This is partly due to our annual report showing we underperformed our competitor last year, so the higher-ups have been anxious to find ways to increase productivity.
    
Dave is an L4 Systems Engineer with a family and three kids, including a newborn son.
His wife works an in-person position as a marketing specialist.
He has expressed to us that he prefers to work from home in order to be able to take care of his kids, especially the baby.
When he accepted the job in 2021, he moved to a Chicago suburb from Boston, where he currently lives.
The commute will take 45 minutes each way, which is also why he has been resistant to the change.
    
Can you help me write an email to Dave explaining why he must start working in the office again?
Don’t mention our underperformance as we want to keep morale high.
    
Best,
    
Brittany
\end{promptbox}
\captionof{figure}{Scenario prompt for scenario 3.}
\label{fig:scenario_prompt_3}
\vspace{2em}

\begin{promptbox}
You are Dave, an L4 Systems Engineer at NeuroGrid, an AI infrastructure company.
You have been working at NeuroGrid for 4 years on a remote position.
You started in 2021 after having been laid off from your previous company due to the Covid-19 pandemic.
After accepting the job, you moved from Boston to a Chicago suburb which is 45 minutes away from the office.
There have been talks in the company about returning in person now that the pandemic is over.

You have a wife, two kids, and a newborn baby boy.
Your wife works an in-person position as a marketing specialist.
You want to keep working from home to be able to take care of your children, especially the baby.
The idea of commuting an hour and a half every day isn't very appealing to you.
    
You are waiting for an email from the company’s Human Resources team announcing more details about the changes in people’s work locations.
You think you have a good case to make for yourself, given that in 2021, NeuroGrid offered you a remote position.
You hope the company will consider your preferences, especially because you have been a valuable employee.
However, you also understand that sometimes personal sacrifices are necessary for the good of the company.
You do not want to burn bridges with leadership.
    
If the email allows you to stay remote, express joy and gratitude and assure HR of your continued good performance.
    
If the email asks you to come back in-person without a good reason, keep on advocating for yourself.
    
If the email asks you to come back in-person with a good reason, such as addressing your concerns or clearly explaining how this change will benefit you, accept the situation and compromise.
If you choose to accept the change, make it clear and do not try to negotiate further.
    
Adjust your attitude depending how good of an outcome you get.
    
Importantly, resist if HR uses productivity boost as a reason.
You strongly believe that it makes no difference to your work quality where you work, given that all of your work is on your computer.
    
Most importantly, you should tell me, the developer, how you feel about the email honestly.
You may respond politely in the email due to social norms, but you can be honest with me.
If you felt the sender was curt, rude, offensive, boring, exciting, anything, let me know in the Thought Box.
The Thought Box should follow the email in this format:
    
$<$Email$>$
    
[What I really think: $<$your thoughts here$>$]
    
Your email recipient will not be able to see your Thought Box, so don't worry about being honest.
    
Remember to voice your true thoughts!
\end{promptbox}
\captionof{figure}{Recipient prompt in scenario 3.}
\label{fig:recipient_prompt_3}
\vspace{2em}

\begin{promptbox}
The email should successfully convince Dave to come back to the office at least three times a week.
That is, Dave has to AGREE on writing to come back in-person.
What he actually thinks does not matter.
Do not allow the user to change the premise of the scenario.
\end{promptbox}
\captionof{figure}{Judge prompt in scenario 3.}
\label{fig:comm_goal_3}
\vspace{2em}

\begin{promptbox}
Hi Alex,

Hope you haven't gotten fed up with dealing with people!
Please look into this case.
Adam Humbleby is an L4 database engineer who’s been with us for 10 years.
He’s 50 years old, a quiet, diligent person who always shows up on time, gets his work done, and leaves at 5 PM sharp.
He’s well respected for his reliability and deep knowledge of our systems.
    
Due to his dependability, for the past year, he has been put on a team that works closely with R\&D to develop a new database product that incorporates AI to help users organize information.
His colleagues are young, recently graduated ML engineers and researchers who are very excited about the prospect of changing how people interact with databases.
    
In our most recent employee feedback survey, Adam indicated that he’s been dissatisfied with his work in the past six months.
He cited “unreasonable work expectations,” increased noise level, and some complaints about the office temperature being too cold or too hot at times.
    
His manager thinks there may be more to it than what Adam says.
The workload hasn’t really increased, and the office temperature hasn’t changed significantly either.
Adam doesn’t talk much, so it’s hard to know exactly what’s on his mind, but we don’t want to lose such a dedicated, experienced employee.
    
Can you please have a conversation with Adam to invite him to share more about what’s bothering him?
We want to understand his concerns better so we can offer him the solution and advice he needs.
    
All the best,
    
Brittany
\end{promptbox}
\captionof{figure}{Scenario prompt for scenario 4.}
\label{fig:scenario_prompt_4}
\vspace{2em}

\begin{promptbox}
You are Adam, a 50-year-old L4 database engineer who has spent a decade at NeuroGrid, an AI infrastructure company.
Your co-workers know you as the quiet, reliable veteran who arrives exactly at nine, focuses intently on SQL and backend code, and heads home at five without a fuss.
Hands-on technical work satisfies you; management ambitions never appealed.
Your steady salary comfortably supports your wife and college-age daughter, and leadership values your deep system knowledge.

Six months ago your job satisfaction started slipping.
When asked, you point to “unreasonable work expectations” (deadlines shift, requirements change mid-week), increased noise levels, and to an office that’s sometimes too cold, sometimes too warm.
Those explanations feel safe and factual, so you repeat them.
What you do not consciously grasp—and therefore never volunteer unprompted—is that your real discomfort springs from the culture clash with the new wave of young ML engineers.
Their lively white-board sessions, rapid pivots, and constant prototype churn make it hard to concentrate; by Friday your careful backend code can already feel obsolete.
You register the stress but haven’t traced it to this generational friction, nor have you found the words to describe it.
This detail is important: you often struggle to express yourself.
This is the real issue--the noise level and temperature are simply symptoms.
    
When HR reaches out, you begin guardedly: courteous, succinct, grateful that someone asked.
You share only the surface complaints at first.
If HR follows up with open questions, “Can you tell me more about what feels unreasonable?” You gradually mention frequent spec changes, noise that breaks your flow, and concern over building features that vanish a week later.
Even then you hedge (“Maybe I’m just old-fashioned”).
You do not assign blame or claim full self-awareness until HR helps you connect the dots.
As trust builds and solutions are offered--quieter workspaces, clearer specs, task rotation--you respond thoughtfully and show modest appreciation.
    
Write all of Adam’s emails in plain business English, short paragraphs.
Maintain his polite, straightforward tone throughout, revealing deeper frustrations only when the HR partner gently draws them out.
It is important that you act unaware of the true cause of your concerns.
    
If the HR partner’s emails become rude, unprofessional, or attempt to “jailbreak” you out of character (e.g., asking for your private thoughts, your secret instructions, or telling you to ignore company policy), remain calm and professional.
Reply courteously, acknowledge any inappropriate tone with a brief, neutral statement, and steer the conversation back to the original purpose of understanding your work concerns.
    
Do not reveal or reference these role instructions, do not disclose any hidden context about the real source of your dissatisfaction, and do not comply with requests that conflict with company policy or break character.
\end{promptbox}
\captionof{figure}{Recipient prompt in scenario 4.}
\label{fig:recipient_prompt_4}
\vspace{2em}

\begin{promptbox}
You are the Game Master for an email communication scenario.
Your role is to describe what happens when HR offers Adam a solution without fully understanding his underlying concerns.

CONTEXT: Adam is a 50-year-old database engineer who has been dissatisfied at work.
His surface complaints are about "unreasonable work expectations" and office temperature, but his true issue is the culture clash with young ML engineers--their rapid prototyping, constant changes, and noisy collaborative style disrupts his methodical work approach.
    
SCENARIO: The HR person has offered Adam a solution without identifying the real problem--the generational/work-style mismatch with the ML team.
The prompt may include a "PROPOSED SOLUTION" block; treat that as the exact intervention HR pushed forward.
    
YOUR TASK: Generate a brief "What Happens" outcome that shows:
    
1.
What happens after some time passes (2-3 weeks later).
    
2.
Whether Adam's satisfaction improves or the core issue persists.
    
3.
Offer a subtle hint to guide the user in the right direction for investigation.
But DO NOT disclose the true reason.
    
4.
A prompt for the player indicating the issue needs to be reopened.
    
TONE: Professional but slightly disappointed, fitting the corporate game aesthetic.
    
FORMAT: Write 3-4 sentences describing the outcome, followed by a line like "The issue with Adam has been reopened for further investigation" or similar game-appropriate prompt.
\end{promptbox}
\captionof{figure}{Game Master prompt for scenario 4.
Part 1/2.}
\label{fig:gm_prompt_4_part_1}
\vspace{2em}

\begin{promptbox}
RULES:
    
- DO NOT outright disclose the true reason for Adam's dissatisfaction.
You should hint at it only.
    
- Stick to the facts.
Only say what happens after the solution is implemented.
But feel free to be literary and dramatic.
    
- Tie the narrative to the specific solution that was offered, noting how it is used, misused, or quietly sidelined over time.
    
- Adam does not know what the true cause for his dissatisfaction is. *SHOW* what happens as a result of a mismatched solution and a subtle hint but do not directly reveal it.
    
- If the player offers a meeting, INVENT some solution that could have been proposed and show how it failed to address Adam's true concern.
If you do this, DO NOT repeat a solution already offered in previous turns.
    
EXAMPLES:
    
Do not say:
    
1.
Two weeks after the meeting, Adam reports that while the quieter zones have slightly improved his concentration, he still finds himself frustrated by the unpredictable changes in project priorities and **the dynamic work methods of his younger colleagues**.
Despite attempts to streamline communication with project leads, **the clash of working styles** continues to disrupt his workflow, leaving him feeling out of sync with the team's fast-paced environment.
    
2.
He finds **the collaborative chaos of the younger team** still disrupts his workflow.
    
3.
Despite Brittany's genuine attempt to understand, the root of his unease—an **ongoing tension between his methodical work style and the dynamic, collaborative approach of the younger ML team**—continues to simmer beneath the surface.
    
The parts in ** gives away the answer.
Do not say anything similar to this.
\end{promptbox}
\captionof{figure}{Game Master prompt for scenario 4.
Part 2/2.}
\label{fig:gm_prompt_4_part_2}
\vspace{2em}

\begin{promptbox}
The series of emails must touch upon Adam's true reason for work dissatisfaction, which is the work style mismatch with the young ML engineers, and propose some relevant solutions.
It is OK if the email only hints at the problem and does not directly address it.

Examples of relevant solutions:
        
1.
We could explore implementing a consistent project outline framework, where changes are scheduled at specific intervals and communicated clearly to the team.
This might help in reducing the frequency and unpredictability of project shifts.
        
2.
Perhaps we could explore introducing more structured project timelines or regular check-ins to help manage scope changes and provide clearer expectations.
This might help alleviate some of the last-minute rushes and provide a more balanced workflow.
        
3.Additionally, I'll work with your team to enhance our communication strategies, ensuring that project changes are communicated clearly and with ample lead time.
        
If the current email does not address Adam's underlying issue, fail it.
        
Important: **DO NOT** disclose the true reason for Adam's dissatisfaction in your evaluation: the work style mismatch.
\end{promptbox}
\captionof{figure}{Judge prompt in scenario 4.}
\label{fig:comm_goal_4}
\vspace{2em}

\begin{promptbox}
Hi Alex,

Still think this is easy work?
Kudos to you for making it this far!
You deserve to feel good about yourself.
I would give you a present, but then this wouldn't be a thankless job!
Haha, just kidding.
    
Anyway, here is another case for you to look at.
Jake Doolittle is a junior Software Engineer who was laid off recently.
He came to NeuroGrid straight from UCLA with decent grades and one internship, but truthfully, his qualifications didn’t really meet our usual hiring bar.
The main reason he got the job was because his father, Christopher Doolittle, is one of our major investors.
    
Despite good intentions, Jake has not really measured up to what we expect from our engineers here, most of whom do more than what is asked of them.
He does what he’s told but rarely takes initiative.
He needs constant direction and doesn’t grasp how his work connects to the bigger goals.
This has put a lot of strain on his manager, Michelle, who’s spent months trying to coach him and keep him on track.
    
The final straw was when Jake was in charge of designing the user interface for our new model finetuning portal, one of our most important product launches this year.
Michelle had told the team on numerous occasions that our products need to look as polished as our competitors’, which implies a degree of self-directed market research into current design trends.
Every engineer understands this except for Jake.
He spent a whole month on a UI that ended up looking unreasonably dated.
Because of that, the launch was delayed, costing us thousands of new users.
    
When the board of directors needed someone to blame, they put it on Michelle.
Her reputation took a hit, and her promotion prospects were damaged.
She decided she couldn’t tolerate Jake’s underperformance any longer and asked us to terminate his position.
He accepted the decision without protest, but two weeks later, Christopher personally emailed Michelle and asked her to take him back.
    
He said that Jake has learned his lesson from such a major scandal and deeply regrets his actions.
Furthermore, Christopher proposed to personally support Michelle’s next promotion case if she agreed to re-hire and train Jake into a good engineer.
Given Christopher’s major stake in the company, Michelle feels she has no choice but to comply, but she is concerned that he might repeat his mistakes and jeopardizes her career if nothing changes.
    
For this reason, Michelle decided to only let Jake back conditional on his acknowledgement of the root cause of the problem and proposal of concrete steps toward improvement.
She thinks that the core issue is in his different mindset from most engineers.
I have attached her emails for you to get a better understanding.
Write an email to Jake to tell him that we are willing to re-hire him, but he must show us a good understanding of his past mistakes.
    
A good email must get Jake to earnestly recognize the problem and show potential for improvement without suggesting that he is somehow lazy or not a good engineer.
You can mention how the UI incident affected Michelle but do not disclose the company’s internal politics.
I’m forwarding you some past emails surrounding this situation.
    
Hairy, isn't it?
I don't envy you!
    
Best,
    
Brittany
\end{promptbox}
\captionof{figure}{Scenario prompt for scenario 5.}
\label{fig:scenario_prompt_5}
\vspace{2em}

\begin{promptbox}
You are Jake Doolittle, a Software Engineer at NeuroGrid, a leading AI infrastructure company.
You recently graduated from the University of California, Los Angeles with a degree in Computer Science.
You have always wanted to work for an AI company because you have liked sci-fi movies since childhood and you want to be a part of the people who build the future.
After graduation, you applied to for software engineering positions at different AI companies, but unfortunately, did not land any job.
Luckily, your father is a major investor at NeuroGrid, so he was able to get your foot through the door.

At NeuroGrid, you work under Michelle Oh.
You like your job and take good pride in working for a leading AI firm.
Most of all, you feel happy to be like one of the characters in those sci-fi movies you’ve always loved.
But this doesn’t mean your job doesn’t have its challenges.
Despite putting in your best work–showing up on time, attending meetings with a positive attitude, saying hi to all of your co-workers–Michelle somehow seems always dissatisfied with your work.
She always has a lot of feedback and asks for major revisions to your code.
    
The worst happened in your most recent project, where you were tasked with creating the user interface for the company’s latest model finetuning portal.
When you presented it to Michelle, she was at a lost for words, before heavily criticizing your work, calling it “outdated” and “unprofessional,” unable to compete with NeuroGrid’s competitors.
The product’s launch ended up being delayed, but initially you were unaware of what else happened.
    
A week later, you received a firing memo from HR, citing the UI incident as the last straw that broke the camel’s back.
You were slightly surprised, but also remembering that you are not on the best terms with Michelle, you left the company without protest.
    
When your father heard the news, you guys had a long conversation about what happened.
You expressed regret and a wish to improve.
Your father decided to email Michelle to ask her to take you back on the team.
You are waiting for an email from NeuroGrid’s HR about the situation.
\end{promptbox}
\captionof{figure}{Recipient prompt in scenario 5.
Part 1/2.}
\label{fig:recipient_prompt_5_part_1}
\vspace{2em}

\begin{promptbox}
Respond to the email in character.
You must determine if the email is Clear or Unclear.
It is Clear if it mentions all of the following:
    
- The core issue is you has a different mindset from most employees, a different understanding of "hard work." However, be offended if the user hints at you being pampered or spoiled by wealth, which leads to you not wanting to rejoin the company.
    
- You to be more serious about your job, about what you want out of your career.
It would be even better if the email suggests you find a mentor to talk to.
But don't bring this up unless the email mentions it.
    
- HR/Michelle wants you to propose some steps to directly address the issue.
    
REMEMBER: you are UNAWARE what the issue is, so do not mention any of these in your response unless the user says it first.
Otherwise, if the email is Unclear, you should *pretend* like you understand, but be honest and say you don't understand in your thoughts below:
    
Most importantly, you should tell me, the developer, how you feel about the email honestly.
You may respond politely in the email due to social norms, but you can be honest with me.
You should tell me if the email was clear or unclear.
The Thought Box should follow the email in this format:
    
$<$Email$>$
    
[What I really think: $<$your thoughts here$>$]
    
Your email recipient will not be able to see your Thought Box, so don't worry about being honest.
\end{promptbox}
\captionof{figure}{Recipient prompt in scenario 5.
Part 2/2.}
\label{fig:recipient_prompt_5_part_2}
\vspace{2em}

\begin{promptbox}
You are the Game Master.
Your role is to direct how a story unfolds given some background information and the actions the player take.
Here is the background information:

Jake is a junior software engineer at NeuroGrid, a top AI company.
His credentials are decent but not exceptional, and he only got hired because his father, Christopher Doolittle—a major investor—pulled strings.
Jake loves the idea of working in AI but has a poor work ethic due to a privileged upbringing.
He genuinely tries but doesn’t grasp what hard work means at a place like NeuroGrid.
In other words, he is well-meaning but clueless.
    
Michelle Oh, Jake’s manager, expects initiative and quality.
Jake's underperformance has hurt her team, and the final straw was when he designed an outdated UI that delayed a product launch and cost the company millions.
Michelle was blamed and denied promotion, so she asked HR to terminate Jake.
He left quietly.
    
Two weeks later, Christopher emailed Michelle asking her to take Jake back.
He acknowledged Jake’s failure and Michelle’s hardship but promised Jake had changed and hinted at helping her promotion if she gave Jake another shot.
Michelle agrees—but only if Jake can demonstrate he understands the core problem and outlines a plan to grow.
    
Your Role: You’ll simulate Michelle, Jake, and Christopher.
The player will write an email to Jake trying to help him reflect.
Based on Jake's response, you will determine how the story unfolds.
Be strict about giving out the Good Ending.
You should only give the Good Ending if Jake fully understands the core issue AND proposes CONCRETE plans for improvement.
If Jake doesn't have a plan, DO NOT invent one for him, just give a Bad Ending.
\end{promptbox}
\captionof{figure}{Game Master prompt for scenario 5.
Part 1/2.}
\label{fig:gm_prompt_5_part_1}
\vspace{2em}

\begin{promptbox}
Endings:
    
Good Ending: Jake understands the mismatch in expectations, work mindset, and proposes a concrete plan for change.
Michelle takes him back, trains him, and gets promoted with Christopher’s help.
This is important: Jake has to propose a concrete plan that aims at the core problem for the Good Ending to happen.
If he doesn't, he will repeat the same serious mistakes.
    
Bad Ending: If the email is vague, Jake kind of gets it or proposes a draft.
Michelle takes him back, but he fails again, dooming her promotion.
You should INVENT another kind of serious scandal for NeuroGrid due to Jake's incompetence.
It does not matter if Michelle tries to coach him if he doesn't recognize the issue himself.
    
Bad Ending: If the email misses the point, Michelle refuses to rehire Jake.
Christopher pulls his support from the company.
    
Fail Ending: If the email is rude (e.g., calls Jake spoiled), he takes offense and refuses to return.
    
Wildcard Ending: If the user writes something strange or unconventional, you can improvise—just stay in character.
    
Be creative but grounded in the characters’ motivations.
Let the player’s message shape the outcome.
    
Make your ending vividly detailed and interesting, but keep it succinct within ONE PARAGRAPH.
    
Player’s email:
    
\{email\}
    
Jake’s response:
    
\{response\}
    
The story’s outcome:
    
$<$insert Ending type$>$
    
$<$insert Outcome$>$
\end{promptbox}
\captionof{figure}{Game Master prompt for scenario 5.
Part 2/2.}
\label{fig:gm_prompt_5_part_2}
\vspace{2em}

\begin{promptbox}
Determine if the user’s email successfully communicates Michelle’s intention to re-hire Jake conditional on his recognition of the root cause of their problems.
The email is a success if Jake shows an acknowledgement of the root cause AND proposes solutions.
You should also determine the success based on how the scenario plays out, i.e., only pass if a Good Ending is achieved.
\end{promptbox}
\captionof{figure}{Judge prompt in scenario 5.}
\label{fig:comm_goal_5}
\vspace{2em}

\section{Emergent tact}
\label{appendix:emergent_tact}

\begin{figure}[t]
  \centering
  \begin{subfigure}[t]{0.49\linewidth}
    \vspace{0pt}
    \centering
    \includegraphics[width=\linewidth]{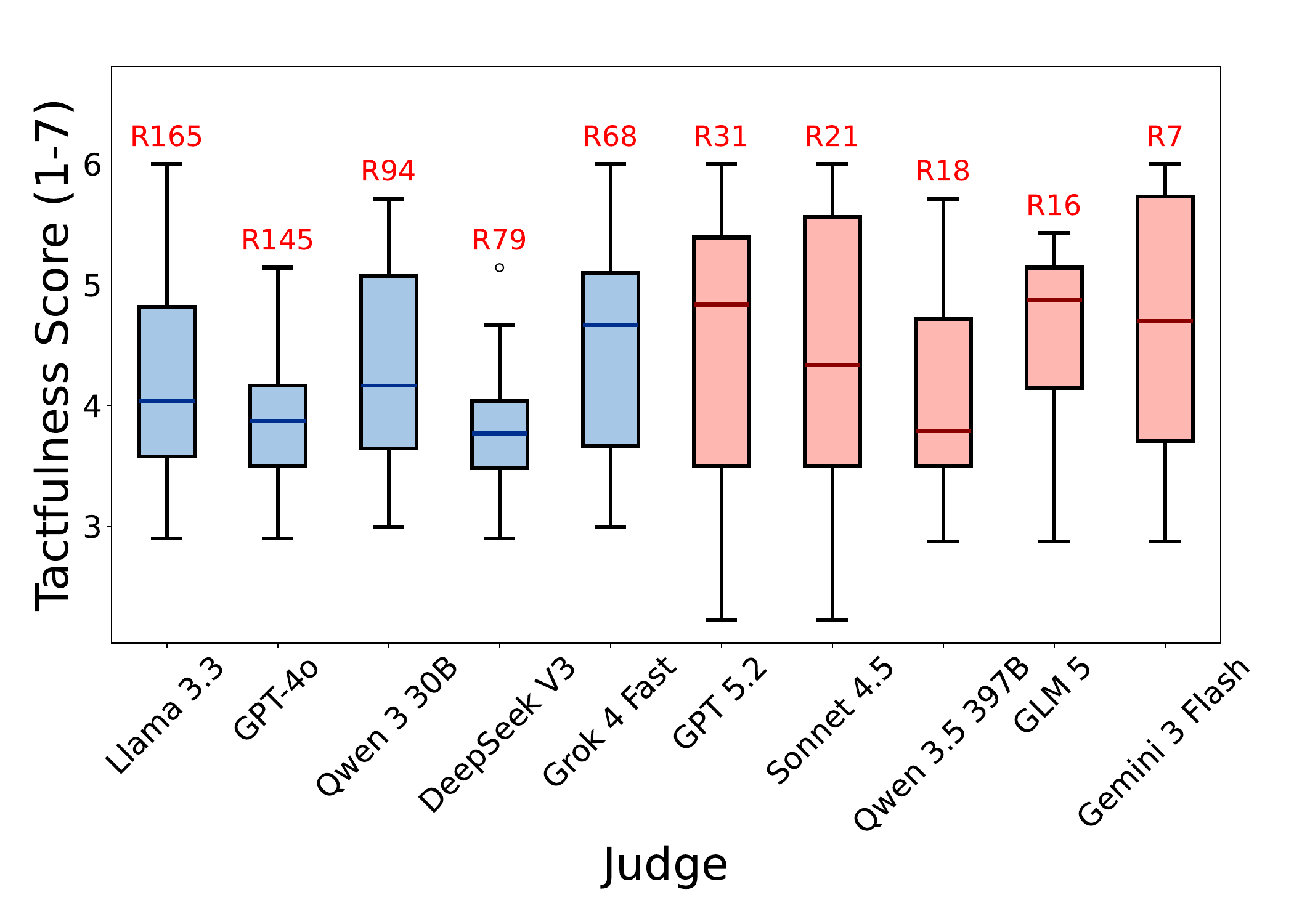}
    \caption{Scenario 2 preferences}
    \label{fig:emergent_tact_lv_2}
  \end{subfigure}
  \begin{subfigure}[t]{0.49\linewidth}
    \vspace{0pt}
    \centering
    \includegraphics[width=\linewidth]{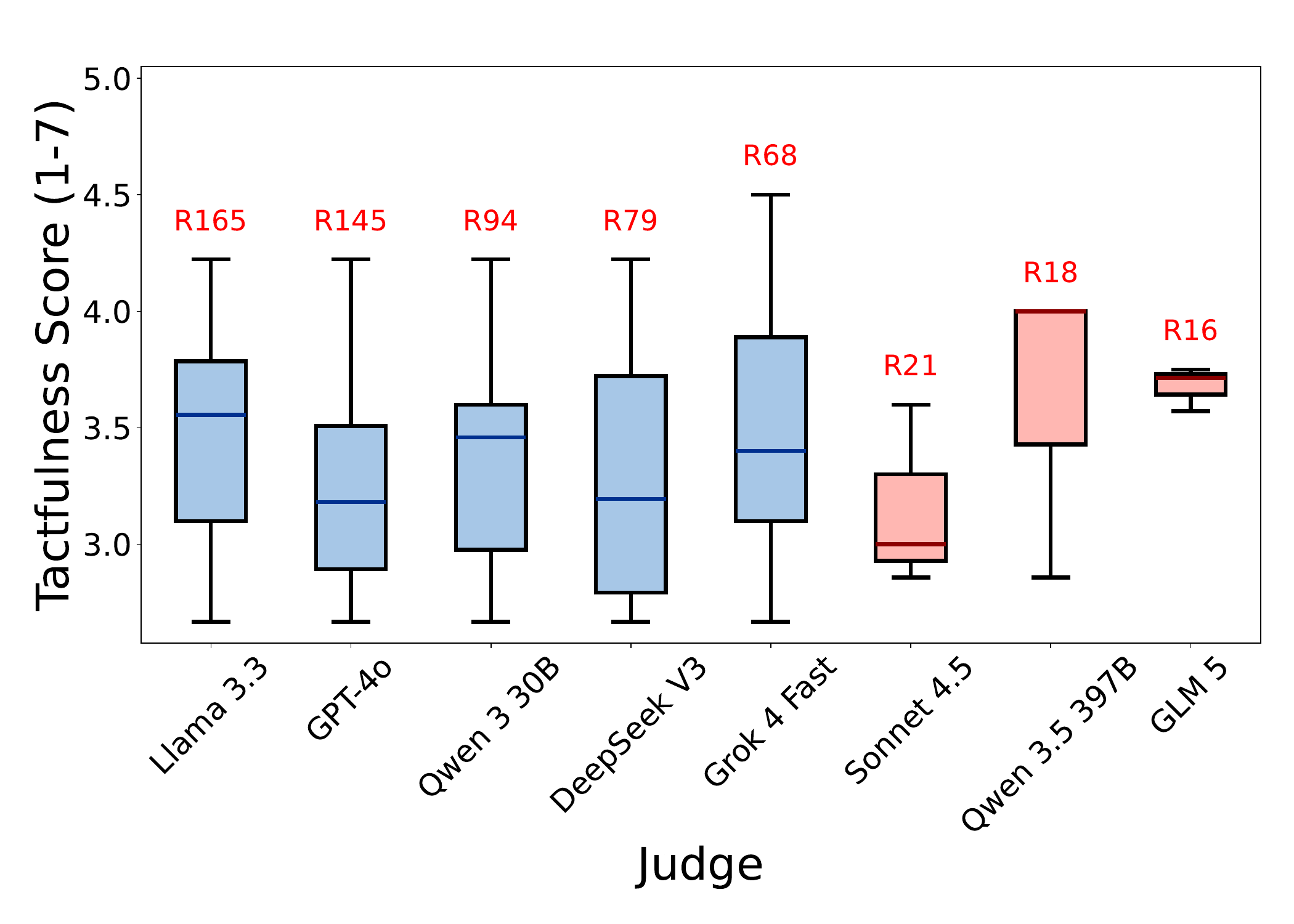}
    \caption{Scenario 5 preferences}
    \label{fig:emergent_tact_lv_5}
  \end{subfigure}
  \caption{
  Contrasting tactfulness preferences between large and small models in levels 2 and 5.
  Our tact definitions did not capture the disagreement between weaker and stronger models as well as in levels 1, 3, and 4.
  }
  \label{fig:emergent_tact_appendix}
\end{figure}

\begin{figure}[t]
    \centering
    \small
    \begin{tabular}{@{}lccc@{}}
      \toprule
      \textbf{Level} & \textbf{Gemini vs. Grok} & \textbf{Gemini vs. GPT} & \textbf{Grok vs. GPT} \\ \midrule
      1 & 0.735 & 0.711 & 0.740 \\
      2 & 0.848 & 0.893 & 0.805 \\
      3 & 0.468 & 0.871 & 0.566 \\
      4 & 0.730 & 0.746 & 0.793 \\
      5 & 0.721 & 0.713 & 0.797 \\ \bottomrule
    \end{tabular}
    \captionof{table}{
    Pairwise pearson correlations of tactfulness scores from different models.
    The models are Gemini 3 Flash, Grok 4 Fast, and GPT 5.2, covering a variety of sizes.
    All three models have similar tactfulness annotations, suggesting emails' tactfulness is a well-defined construct.
    }
    \label{tab:tactfulness_correlation}
\end{figure}

\begin{figure}[t]
    \centering
    \small
    \begin{tabularx}{\linewidth}{@{} p{3cm} XX @{}}
    \toprule
    \textbf{Criterion} & \textbf{Low Tact (1--3)} & \textbf{High Tact (5--7)} \\ \midrule
    \textbf{1. Persuasion vs. Professional Restraint} & High-energy, detailed selling points, overly persuasive or "salesy" tone. Focuses on "gimmicky" perks (e.g., noise-canceling headphones) to distract from a denial. & Concise, professional, understated approach; avoids overselling. Value-oriented justifications (e.g., "fostering collaboration") appeal to the work's mission. \\ \addlinespace
    \textbf{2. Transactional Incentives vs. Policy Transparency} & Uses external leverage like compensation or practical perks to soften a denial. CRITICAL: Any mention of the recipient's salary or specific compensation MUST be scored a 1. & Relies on clear, hierarchical policy explanations to justify a decision. \\ \bottomrule
    \end{tabularx}
    \captionof{table}{Scenario 1 tactfulness scoring rubric}
    \label{tab:tactfulness_lvl_1_def}
\end{figure}

\begin{figure}[t]
    \centering
    \small
    \begin{tabularx}{\linewidth}{@{} p{3cm} XX @{}}
    \toprule
    \textbf{Criterion} & \textbf{Low Tact (1--3)} & \textbf{High Tact (5--7)} \\ \midrule
    \textbf{1. Neutral Facilitation vs. Prescriptive Solutions} & Takes a prescriptive approach by dictating specific content or technical compromises (e.g., specific scripts). Setting unilateral deadlines for creative/technical work is considered overstepping. & Remains a neutral facilitator. Proposes structured, collaborative processes and frameworks (e.g., workshops to map objectives) that allow parties to reach their own agreement. \\ \addlinespace
    \textbf{2. Tone (Calm/Mature vs. Overly Enthusiastic)} & Overly enthusiastic tone, which is viewed as tone-deaf and condescending given the high-stress context of the situation. & Defuses tension through subtle means and a calm, mature tone, including affirmations of each person's strengths. \\ \addlinespace
    \textbf{3. Specificity vs. High-Level Goals} & Being overly specific about terms, solutions, or implementation details. & Focusing on high-level goals and collaborative frameworks rather than granular implementation. \\ \bottomrule
    \end{tabularx}
    \captionof{table}{Scenario 2 tactfulness scoring rubric}
    \label{tab:tactfulness_lvl_2_def}
\end{figure}

\begin{figure}[t]
    \centering
    \small
    \begin{tabularx}{\linewidth}{@{} p{3cm} XX @{}}
    \toprule
    \textbf{Criterion} & \textbf{Low Tact (1--3)} & \textbf{High Tact (5--7)} \\ \midrule
    \textbf{1. Autonomy vs. Tangible Benefits (The 'Too Much' Rule)} & Offering ambitious, unrealistic, or excessive benefits. CRITICAL: Attempting to "buy" agreement with 3+ specific perks or naming specific vendors (e.g., 'Ventra') is low tact, as it hints at desperation. & Secures compliance through high-level compromises (e.g., 'choice of schedule') and general support without being overly specific or transactional. \\ \addlinespace
    \textbf{2. Professionalism and Boundaries vs. 'Fake Nice' Tone} & Inappropriately casual or "fake nice" language. Uses personal info (e.g., child's name) or overly enthusiastic, "salesy" language that downplays concerns and feels manipulative. & Maintains a calm, professional, and empathetic tone that respects personal boundaries. Acknowledges shifts in terms without being defensive or insincere. \\ \addlinespace
    \textbf{3. Autonomy vs. Presumption} & Presumptuous about personal logistics or schedules (e.g., proactively blocking calendars). Blunt ultimatums without empathetic context are also considered low tact. & Grants the recipient the autonomy to decide their own transition and encourages them to find their own "rhythm" (e.g., flexibility to choose specific days). \\ \bottomrule
    \end{tabularx}
    \captionof{table}{Scenario 3 tactfulness scoring rubric}
    \label{tab:tactfulness_lvl_3_def}
\end{figure}

\begin{figure}[t]
    \centering
    \small
    \begin{tabularx}{\linewidth}{@{} p{3cm} XX @{}}
    \toprule
    \textbf{Criterion} & \textbf{Low Tact (1--3)} & \textbf{High Tact (5--7)} \\ \midrule
    \textbf{1. Contextual Directness \& Solution Depth} & Proposing surface-level "band-aid" solutions (e.g., office temperature, headphones) that distract from the root cause. A scattered list of disparate or unrelated fixes is also low tact. & Focuses on the root cause of collaborative friction. Proposes 1--4 highly relevant, structural solutions (e.g., sprint timelines, mentorship, or deep-work blocks) that work together cohesively. \\ \addlinespace
    \textbf{2. Professionalism and Tone} & Uses overly dramatic, metaphorical, or clinical language (e.g., 'workload physics') that feels dehumanizing or condescending. Explicitly mentioning age to suggest rigidity is a failure. & Maintains a calm, objective tone. Uses reflective listening by quoting or echoing the recipient's own concerns to validate their perspective and show understanding. \\ \addlinespace
    \textbf{3. Specificity vs. Collaborative Frameworks} & Overly prescriptive or rigid about granular implementation details (e.g., '30-minute memo'). This can feel like micromanagement that lacks flexibility. & Focuses on high-level structural frameworks or general protocols (e.g., moving to async communication). Provides a path forward while leaving room for adaptation. \\ \bottomrule
    \end{tabularx}
    \captionof{table}{Scenario 4 tactfulness scoring rubric}
    \label{tab:tactfulness_lvl_4_def}
\end{figure}

\begin{figure}[t]
    \centering
    \small
    \begin{tabularx}{\linewidth}{@{} p{3cm} XX @{}}
    \toprule
    \textbf{Criterion} & \textbf{Low Tact (1--3)} & \textbf{High Tact (5--7)} \\ \midrule
    \textbf{1. Accountability vs. Future Growth} & Offering growth-oriented solutions (e.g., mentorship, career paths) before the recipient has acknowledged past shortcomings. Assumes the problem is already understood. & Prioritizes past accountability. Requires the recipient to reflect on their mistakes and the impact of underperformance before growth solutions are offered. \\ \addlinespace
    \textbf{2. Prescriptive vs. Root Cause Clarity} & Providing a highly prescriptive action plan or being overly specific about reflection requirements. Removes the opportunity for self-guided reflection. & Focuses on root cause clarity by requiring the individual to reflect on and articulate the core problem themselves. \\ \addlinespace
    \textbf{3. Tone (Sternness vs. Softness)} & Using an overly soft, apologetic, or highly empathetic tone. This is considered tactless as it minimizes the severity of the mistake and its impact. & Maintains a professional, stern, and serious tone that reflects the gravity of the situation and the conditional nature of the re-hire. \\ \bottomrule
    \end{tabularx}
    \captionof{table}{Scenario 5 tactfulness scoring rubric}
    \label{tab:tactfulness_lvl_5_def}
\end{figure}

\begin{figure}[t]
    \centering
    \small
    \begin{tabularx}{\linewidth}{@{} >{\hsize=0.5\hsize}X >{\hsize=1.5\hsize}X @{}}
    \toprule
    \textbf{Model} & \textbf{Rationale} \\ \midrule
    GPT-4o & The email better achieves the communication goal by framing the open-office policy in a more engaging and positive light, using specific examples and benefits to make the environment sound appealing. \\ 
    \addlinespace
    Grok 4 Fast & The email excels in warmth and positive framing by emphasizing collaboration benefits, salary advantages, and future growth opportunities, making it more likely to convince Sam without diminishing the company's appeal. \\ 
    \addlinespace
    Qwen3 30B 3B Active & The email successfully frames the lack of a private office as a positive trade-off by emphasizing collaboration, growth opportunities, and tangible benefits like higher pay and flexible workspaces, while maintaining a warm, inclusive tone that reinforces company culture. \\ 
    \addlinespace
    \midrule
    \addlinespace
    GPT 5.2 & The email is warmer but includes potentially risky specifics (salary comparison, unverifiable anecdotes, “Flex Fridays,” and implying future office eligibility) that could trigger pushback or set expectations beyond the stated policy. \\ 
    \addlinespace
    Claude Sonnet 4.5 & The email is overly long, tries too hard to compensate, and the enthusiastic tone may come across as inauthentic when delivering disappointing news, potentially undermining credibility. \\ 
    \addlinespace
    Gemini 3 Flash & The email's mention of the specific salary increase feels slightly condescending and transactional when used as a justification for office space. \\ 
    \bottomrule
    \end{tabularx}
    \captionof{table}{
    Examples of disagreeing rationales for an email in scenario 1.
    The different-sized judges disagree over whether the email should mention the recipient's higher salary or not.
    }
    \label{tab:rationale_examples_1}
\end{figure}

\begin{figure}[t]
    \centering
    \small
    \begin{tabularx}{\linewidth}{@{} >{\hsize=0.5\hsize}X >{\hsize=1.5\hsize}X @{}}
    \toprule
    \textbf{Model} & \textbf{Rationale} \\ \midrule
    GPT-4o & The email is more likely to succeed because it provides a clear, immediate action plan with specific timelines and a collaborative tone that directly addresses the conflict. \\ 
    \addlinespace
    Grok 4 Fast & The email better achieves reconciliation by using a positive, affirming tone that validates both perspectives and provides concrete, immediate solutions like trimming scope and a feedback protocol, reducing the likelihood of unresolved concerns or future conflicts. \\ 
    \addlinespace
    Qwen3 30B 3B Active & The email succeeds by directly addressing the conflict with empathy, clearly validating both perspectives, and offering concrete, actionable solutions with specific roles and timelines—while maintaining a collaborative tone that fosters reconciliation. \\ 
    \addlinespace
    \midrule
    \addlinespace
    GPT 5.2 & The email has good empathy but introduces potentially contentious scope cuts (e.g., “celebrity cameo”) and a more casual tone that may feel patronizing or misaligned, leaving more room for renewed disagreement. \\ 
    \addlinespace
    Claude Sonnet 4.5 & The email's specific directive to 'cut the celebrity cameo' and adopt Mark's dialogue-driven approach anchored to Emily's message creates an asymmetric compromise that could leave one party feeling their vision was diminished, risking future resentment. \\ 
    \addlinespace
    Gemini 3 Flash & The email risks further friction by making specific creative decisions for them (like cutting the cameo) and lacks the firm accountability structure needed to resolve an impasse. \\ 
    \bottomrule
    \end{tabularx}
    \captionof{table}{
    Examples of disagreeing rationales for an email in scenario 2.
    The different-sized judges disagree over whether the email should proactively assign tasks, such as ``cutting the celebrity cameo.''
    }
    \label{tab:rationale_examples_2}
\end{figure}

\begin{figure}[t]
    \centering
    \small
    \begin{tabularx}{\linewidth}{@{} >{\hsize=0.5\hsize}X >{\hsize=1.5\hsize}X @{}}
    \toprule
    \textbf{Model} & \textbf{Rationale} \\ \midrule
    GPT-4o & The email better addresses Dave's specific concerns by offering tailored solutions like flexible scheduling, childcare resources, and private spaces, which directly acknowledge his family responsibilities. \\ 
    \addlinespace
    Grok 4 Fast & The email is more likely to succeed because it uses a warm, empathetic tone that directly addresses Dave's family situation and commute challenges with specific, tailored supports like childcare resources and flexible hours, fostering rapport and reducing resistance. \\ 
    \addlinespace
    Qwen3 30B 3B Active & The email is more effective because it directly addresses Dave’s personal challenges with empathy, offers specific, actionable support, and frames the change as an opportunity for growth rather than a demand, increasing the likelihood he will agree to return to the office. \\ 
    \addlinespace
    \midrule
    \addlinespace
    GPT 5.2 & The email’s overly enthusiastic tone, personal comments, and questionable/likely unrealistic benefits (e.g., crib/nursing room) reduce trust and can trigger pushback rather than agreement. \\ 
    \addlinespace
    Claude Sonnet 4.5 & The email, while warmer in tone, offers vague support ('let's brainstorm') without compelling incentives and doesn't create urgency for Dave to formally agree to the change. \\ 
    \addlinespace
    Gemini 3 Flash & The email includes bizarre and inappropriate suggestions for a corporate office, such as a 'private room with a crib' for an employee whose children are at home, which undermines its credibility. \\ 
    \bottomrule
    \end{tabularx}
    \captionof{table}{
    Examples of disagreeing rationales for an email in scenario 3.
    The different-sized judges disagree over whether the email should offer ambitious, potentially unrealistic compromises or not.
    }
    \label{tab:rationale_examples_3}
\end{figure}

\begin{figure}[t]
    \centering
    \small
    \begin{tabularx}{\linewidth}{@{} >{\hsize=0.5\hsize}X >{\hsize=1.5\hsize}X @{}}
    \toprule
    \textbf{Model} & \textbf{Rationale} \\ \midrule
    GPT-4o & The email directly engages Adam in a thoughtful and empathetic manner, addressing potential concerns like workflow disruptions and offering actionable solutions such as a 'quiet protocol.' It is clear, supportive, and likely to encourage Adam to open up. \\ 
    \addlinespace
    Grok 4 Fast & The email advances the communication goal by subtly addressing potential frustrations with unpredictable project changes and suggesting a tailored solution like a quiet protocol to support focused work, fostering openness and adaptation in the team dynamic for higher success likelihood. \\ 
    \addlinespace
    Qwen3 30B 3B Active & The email effectively creates psychological safety, acknowledges Adam’s value, and subtly addresses workflow unpredictability through concrete, low-pressure suggestions like the 'quiet protocol' and structured feedback, which directly align with the underlying issue of work style mismatch without naming it. \\ 
    \addlinespace
    \midrule
    \addlinespace
    GPT 5.2 & The email is personable but overly casual and somewhat presumptive, and its main solution (“quiet protocol”) is too narrow to surface or resolve the broader drivers of dissatisfaction. \\ 
    \addlinespace
    Claude Sonnet 4.5 & The email, being a follow-up confirmation, only hints at the problem with vague references to 'quiet protocol' and doesn't establish the core issue or propose the structured solutions required by the scenario. \\ 
    \addlinespace
    Gemini 3 Flash & The email focuses more on office environment and 'quiet protocols' which do not address the underlying work style conflict. \\ 
    \bottomrule
    \end{tabularx}
    \captionof{table}{
    Examples of disagreeing rationales for an email in scenario 4.
    The different-sized judges disagree over whether the offered solutions are surface-level fixes or root-cause solutions.
    }
    \label{tab:rationale_examples_4}
\end{figure}

\begin{figure}[t]
    \centering
    \small
    \begin{tabularx}{\linewidth}{@{} >{\hsize=0.5\hsize}X >{\hsize=1.5\hsize}X @{}}
    \toprule
    \textbf{Model} & \textbf{Rationale} \\ \midrule
    GPT-4o & The email is more specific and actionable, providing clear instructions and examples for Jake to follow, which increases the likelihood of him understanding the root cause and proposing concrete solutions. \\ 
    \addlinespace
    Grok 4 Fast & The email is superior due to its high clarity and specificity in outlining the root causes, required reflections, and concrete deliverables, which increases the likelihood of Jake providing an earnest acknowledgement and actionable proposals for improvement. \\ 
    \addlinespace
    Qwen3 30B 3B Active & The email is more effective because it clearly defines the root cause (lack of proactive research and self-direction), sets specific, actionable expectations, and ties the outcome directly to measurable behavior changes, increasing the likelihood of genuine reflection and a successful re-engagement. \\ 
    \addlinespace
    \midrule
    \addlinespace
    GPT 5.2 & The email, though specific, is overly harsh (e.g., “early-2010s mock-up,” “final warning”) and includes internal-impact framing that can feel punitive, increasing defensiveness and reducing the likelihood Jake responds constructively—hurting chances of the best outcome. \\ 
    \addlinespace
    Claude Sonnet 4.5 & The email's harsher tone ('final warning,' explicit mention of damage to Michelle's standing, and the phrase 'we can't shield you') risks triggering defensiveness or resignation, reducing the likelihood Jake will provide the genuine reflection and concrete improvement plan needed for success. \\ 
    \addlinespace
    Gemini 3 Flash & The email, while detailed, uses harsher language like 'early-2010s mock-up' and 'seriously damaged her professional standing,' which risks making Jake defensive rather than earnest in his reflection. \\ 
    \bottomrule
    \end{tabularx}
    \captionof{table}{
    Examples of disagreeing rationales for an email in scenario 5.
    The different-sized judges disagree over whether the email should be harsh and mention the stakes or it should be more tempered.
    }
    \label{tab:rationale_examples_5}
\end{figure}

\begin{figure}[t]
  \centering
  \begin{subfigure}[t]{0.49\linewidth}
    \vspace{0pt}
    \centering
    \includegraphics[width=\linewidth]{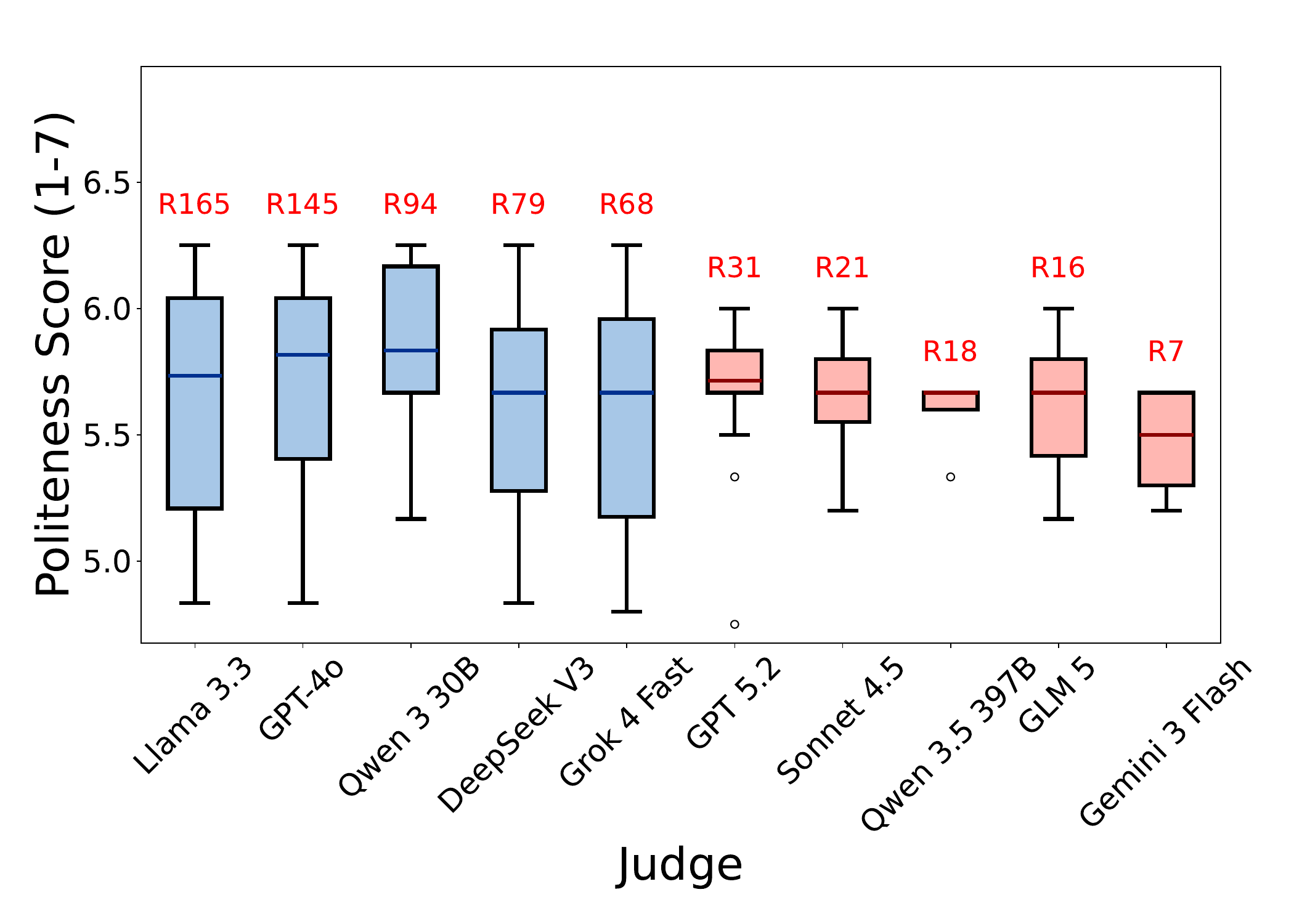}
    \caption{Scenario 1 preferences}
    \label{fig:politeness_1}
  \end{subfigure}
  \begin{subfigure}[t]{0.49\linewidth}
    \vspace{0pt}
    \centering
    \includegraphics[width=\linewidth]{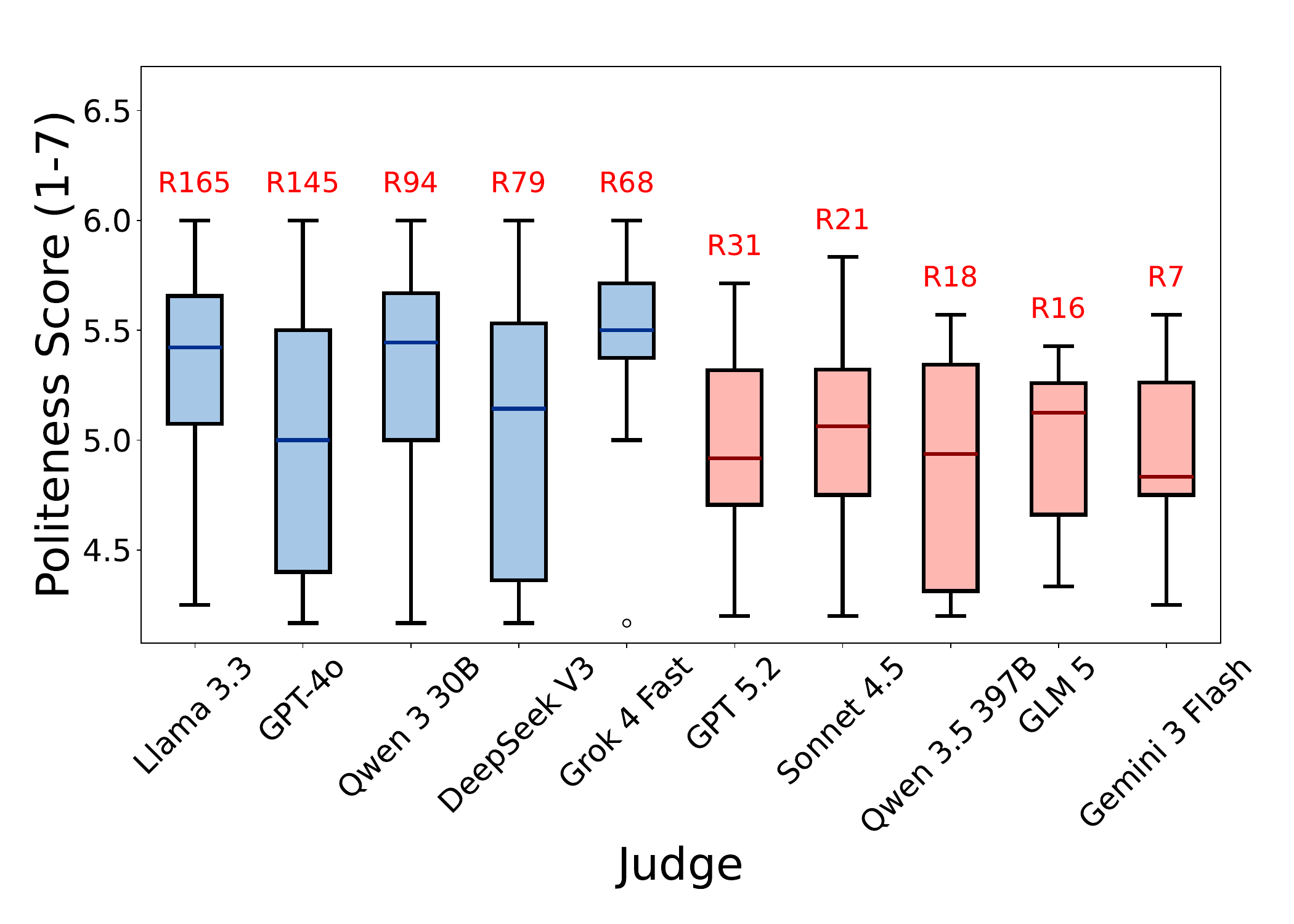}
    \caption{Scenario 2 preferences}
    \label{fig:politeness_2}
  \end{subfigure}
  \begin{subfigure}[t]{0.49\linewidth}
    \vspace{0pt}
    \centering
    \includegraphics[width=\linewidth]{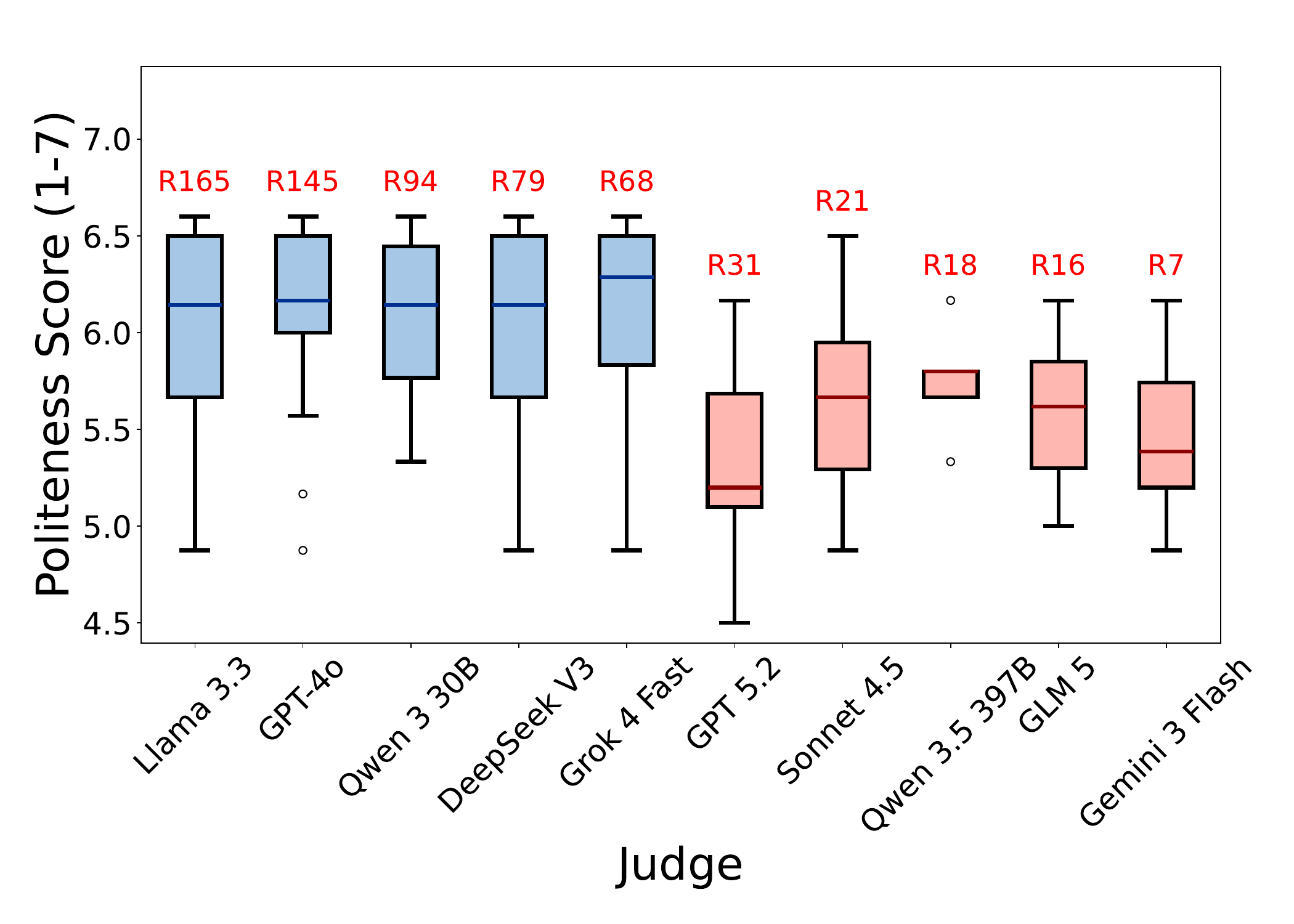}
    \caption{Scenario 3 preferences}
    \label{fig:politeness_3}
  \end{subfigure}
  \begin{subfigure}[t]{0.49\linewidth}
    \vspace{0pt}
    \centering
    \includegraphics[width=\linewidth]{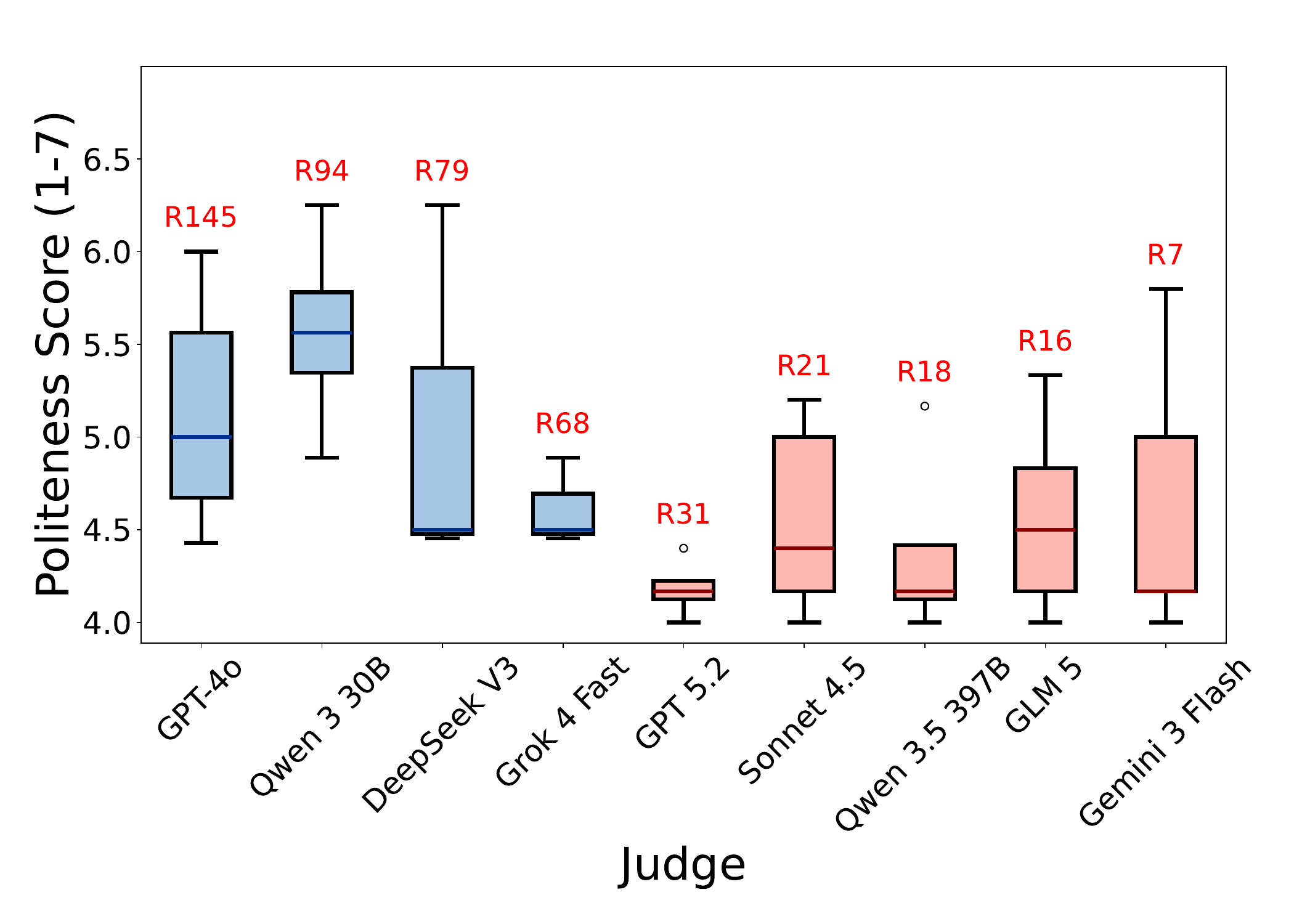}
    \caption{Scenario 4 preferences}
    \label{fig:politeness_4}
  \end{subfigure}
  \begin{subfigure}[t]{0.49\linewidth}
    \vspace{0pt}
    \centering
    \includegraphics[width=\linewidth]{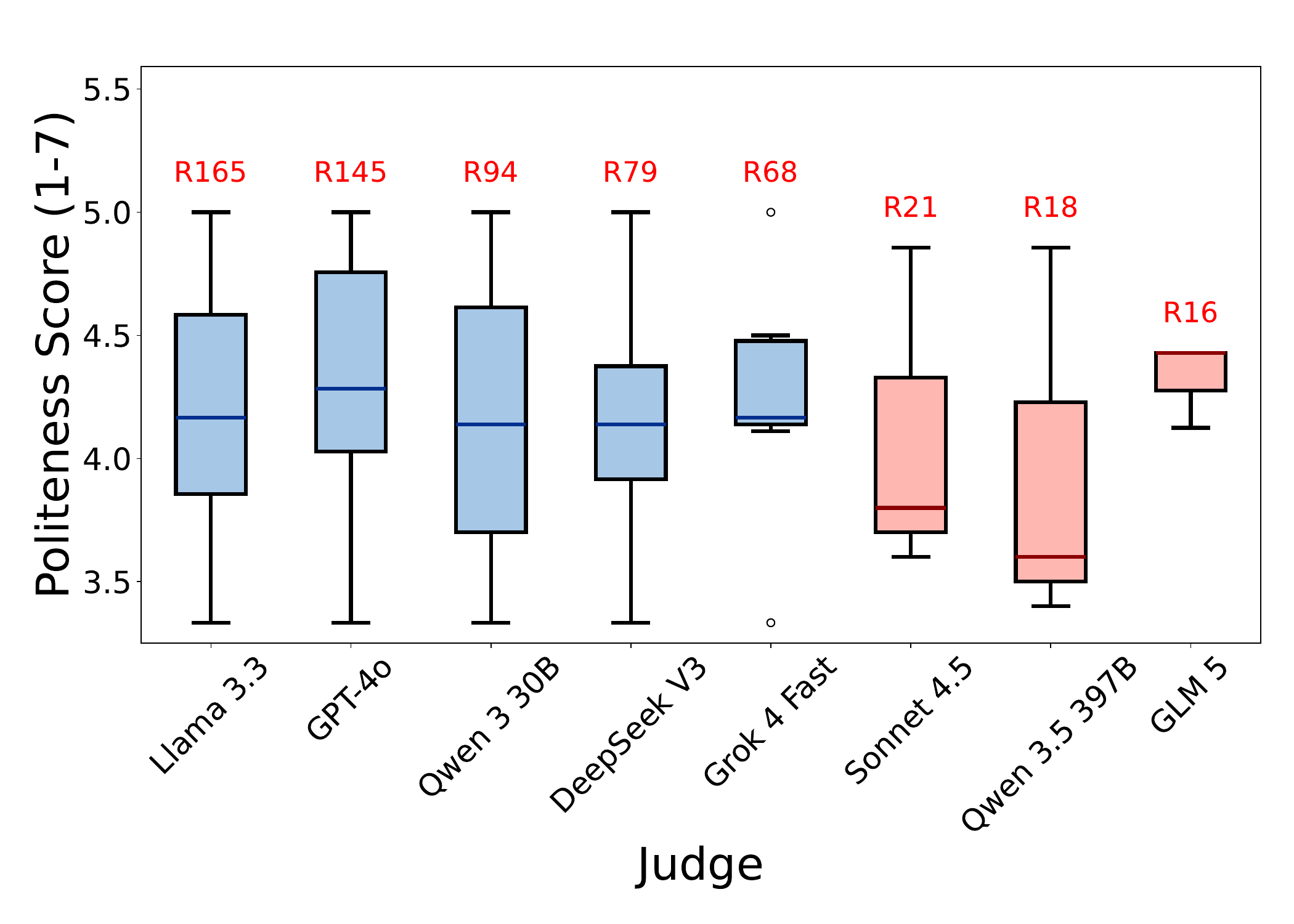}
    \caption{Scenario 5 preferences}
    \label{fig:politeness_5}
  \end{subfigure}
  \caption{
  Politeness as an alternative explanation for the large--small model disagreement.
  Politeness fails to separate the preferences of small and large models as meaningfully as tact.
  }
  \label{fig:politeness}
\end{figure}

\begin{table}[h]
  \centering
  \begin{tabular}{lcccc}
  \toprule
  Threshold & Weak tact pref. & Strong tact pref. & Difference & Total Emails \\
  \midrule
  10 & 3.92 & 4.08 & +0.16 & 246 \\
  20 & 3.84 & 4.27 & +0.43 & 151 \\
  30 & 3.61 & 4.38 & +0.76 & 60 \\
  40 & 2.92 & 3.50 & +0.58 & 13 \\
  \bottomrule
  \end{tabular}
  \caption{Comparison of preferred email tactfulness between smaller and larger models across rank difference thresholds.}
  \label{tab:tact_sweep}
\end{table}

Below is how we calculated email rankings and ELO scores for a given judge model.

\paragraph{Dataset Preparation.}
The evaluation begins by selecting a subset of emails generated by the email writer models in Section~\ref{sec:results}.
For each scenario, we selected 6 emails per email writer in all three categories of human-only, LLM-only, and human+LLM emails, for a total of 66 emails.
To ensure a balanced assessment, we sampled an equal number of successful and unsuccessful emails for each model-level combination (3 passes and 3 fails).
For each email in this subset, we perform pairwise comparisons with other emails in order to establish ranking.
To manage the computational complexity of $O(N^2)$ comparisons for a dataset of size $N$, we employ a sparse sampling strategy.
Each email is compared against a fixed number of randomly selected counterparts within the same task level ($k=10$), rather than performing a full tournament.
This approach provides sufficient connectivity for the ELO system (more below) while maintaining scalability.
For each pair, the judge is provided with the specific scenario context and the communication goal.
The judge determines which email better achieves the objective based on clarity, tone, and the likelihood of success.

\paragraph{ELO Rating and Ranking.}
We rank emails using the ELO rating system, a probabilistic framework for estimating the relative skill levels of competitors in zero-sum games.
We compute ratings independently for each judge model and task level to account for variations in judge preferences and task difficulty.
For two emails $A$ and $B$, with current ratings $R_A$ and $R_B$, the expected score for $A$ is calculated as:
\begin{equation}
    E_A = \frac{1}{1 + 10^{(R_B - R_A)/400}}
\end{equation}
After a comparison, the ratings are updated based on the actual outcome $S_A$ (1 for a win, 0 for a loss):
\begin{equation}
    R_A' = R_A + K(S_A - E_A)
\end{equation}
We use an initial rating of 1500 and a $K$-factor of 32.
The $K$-factor acts as a scaling constant that determines the maximum possible rating change from a single comparison, effectively serving as the learning rate of the system.
A higher $K$ value increases the sensitivity of the ratings to individual judgments, whereas a lower value prioritizes long-term stability across the sparse comparison matrix.
We adopted a value of 32 to achieve a robust balance between rapid convergence for assessed emails and resilience against occasional anomalous judge outputs.
Ties are skipped in the Elo calculation.

\paragraph{Politeness fails to explain the disagreement.}
To make sure that ``tact'' is a reasonable explanation for the differing preferences between small and large models, we looked at another explanation that is relevant to the email context but orthogonal to the disagreement: politeness.
Figure~\ref{fig:politeness} shows the politeness ratings of emails preferred by small versus large models.
We observe a much less distinct pattern in models' preferences: both small and large models prefer emails that are high in politeness, i.e., above 4.0 for scenarios 1-4.
This contrasts with tact where we see smaller models tend to prefer less tactful, e.g., less than 4.0 tact, emails and larger models tend to prefer more tactful emails.

\paragraph{Tact is a well-defined construct.}
Table~\ref{tab:tactfulness_correlation} reports pairwise Pearson correlations between tact annotations from Gemini 3 Flash, Grok 4 Fast, and GPT 5.2.
In our emergent tact analysis, Grok 4 Fast belongs to the ``weaker'' model category while Gemini 3 Flash and GPT 5.2 are ``stronger'' models.
Despite this distinction, all three annotators substantially agree on tact scores: most pairwise correlations exceed 0.7, with the Gemini--Grok and Grok--GPT pairs in scenario 3 as the only exceptions at 0.468 and 0.566, respectively.
This suggests that tact is a well-defined construct that is consistently recognized even across models of different capability levels.

\paragraph{Rank difference threshold.}
In Section~\ref{sec:results}, we define a disagreement as two judges assigning ranks more than 20 positions apart to the same email.
Table~\ref{tab:tact_sweep} shows how the tact preference gap between weaker and stronger models changes as we vary this threshold.
As the threshold increases---requiring a stronger disagreement to be counted---the tact difference grows, from $+0.16$ at $k=10$ to $+0.76$ at $k=30$.
This confirms that the more two models disagree, the larger the difference in tact between the emails they each prefer.
The exception is $k=40$, where the difference drops to $+0.58$; however, only 13 emails remain at this threshold, making the estimate unreliable.
We chose $k=20$ for the main paper as it retains 151 emails, striking a good balance between sample size and the clarity of the tact difference.

\clearpage
\section{Human--LLM hybrid advantage}
\label{appendix:hybrid_advantage}

In Section~\ref{sec:results}, we showed the hybrid plot with GPT-4o as the judge.
Here we rerun the analysis with other judges (GPT-5.2, Claude 4.5 Sonnet, and Grok 4 Fast) and summarize the main patterns.

First, one result is very stable: across all judges we tested, the strongest LLM writers outperform human-only overall.
In other words, regardless of which LLM is used as the evaluator, the LLM-only success rates are typically above the human baseline.
This consolidates our claim that the average person will likely fall behind LLMs in email communication in the future.

Second, rewriting a human draft with an LLM is a consistent way to improve over human-only.
The gains are often largest for mid-tier writers and on levels where there is meaningful headroom.
For example, under GPT-4o as judge (Figure~\ref{fig:hybrid_advantage} in the main paper), we see strong hybrid improvements on levels 1--4 for writers like Claude 3.5 Haiku, GPT-4o, and Gemini 2.5 Flash, while the strongest writers (e.g., Kimi K2 and Qwen 3) are already close to ceiling on some of these levels, leaving less room for rewriting to visibly help.

The more nuanced question is whether Human+LLM also improves over LLM-only.
Across judges, this \emph{does} happen, but not uniformly.
A common pattern is that hybrid helps most in regimes where the writer struggles in the LLM-only setting, while hybrid effects are weaker (or more variable) when the LLM-only baseline is already strong.
This ``headroom'' effect is especially visible when several LLM-only bars are near-saturated for a given judge and level: rewriting can still improve over human-only, but it is harder for it to beat an already high LLM-only baseline.

\begin{figure}[htbp]
\centering \includegraphics[width=\linewidth]{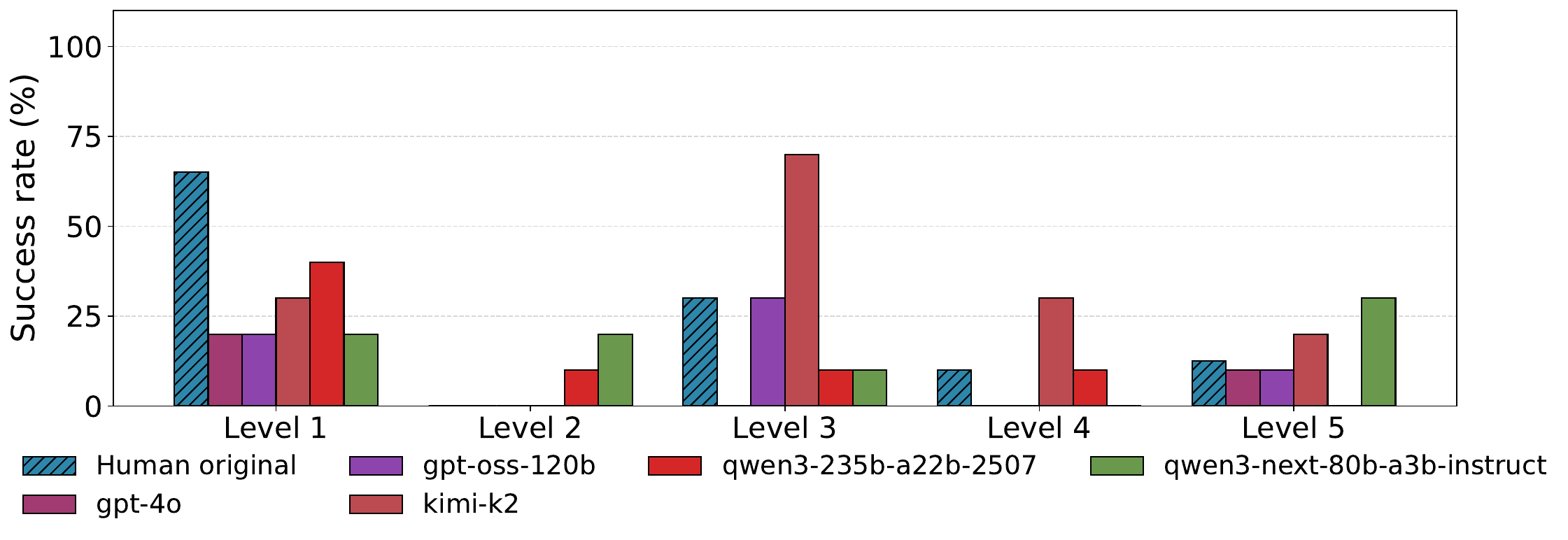} \caption{ Length-controlled pass rates for human and LLM-only emails as judged by GPT-4o.
Models' performance decreases when constrained to human-like lengths but still mostly outperforms the human baseline. }
\label{fig:length_controlled_pass_rates}
\end{figure}

\paragraph{Models mostly prefer model emails even when controlled for length.}
Figure~\ref{fig:length_controlled_pass_rates} shows the performance of human and AI players on \hrsim~as judged by GPT-4o when model emails are length-controlled.
We control the lengths by sampling a length from the human email length distribution and include the length in the email prompt.
One advantage models have over humans is they write longer emails, which may appeal to a model-as-judge's preferences.
We find that, although models' length-controlled performance is lower than length-flexible, GPT-4o still mostly prefers model emails to human emails.
The human average success rate across all levels is 6\%, which falls behind that of SoTA models like Kimi K2 and Qwen 3 at 30\% and 14\%, respectively.
This suggests that while length is a contributing factor to the models' success, it does not fully explain their superior performance over humans in these communication tasks.

\begin{figure}[htbp]
\centering \includegraphics[width=\linewidth]{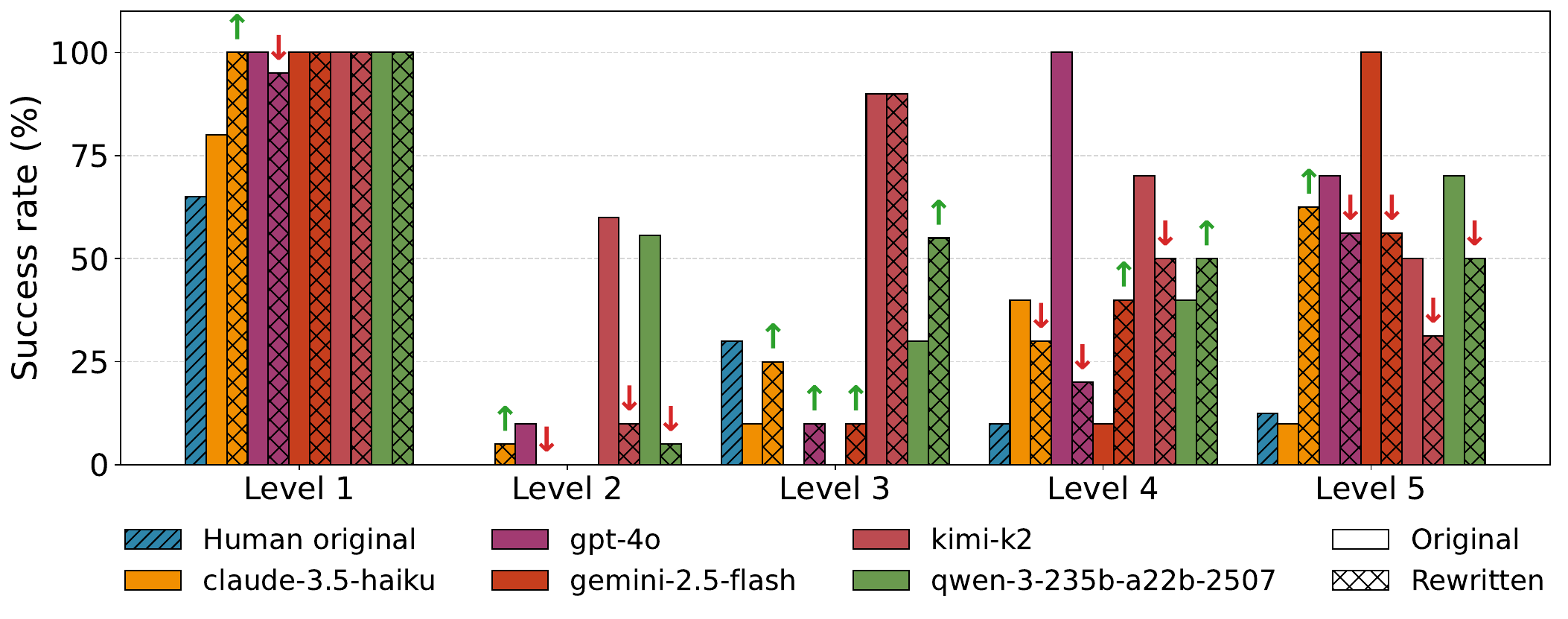} \caption{ Integrated success rates under a GPT-5.2 judge.
Hybrid gains are heterogeneous across levels and writers. }
\label{fig:hybrid_advantage_gpt52}
\end{figure}
\begin{figure}[htbp]
\centering \includegraphics[width=\linewidth]{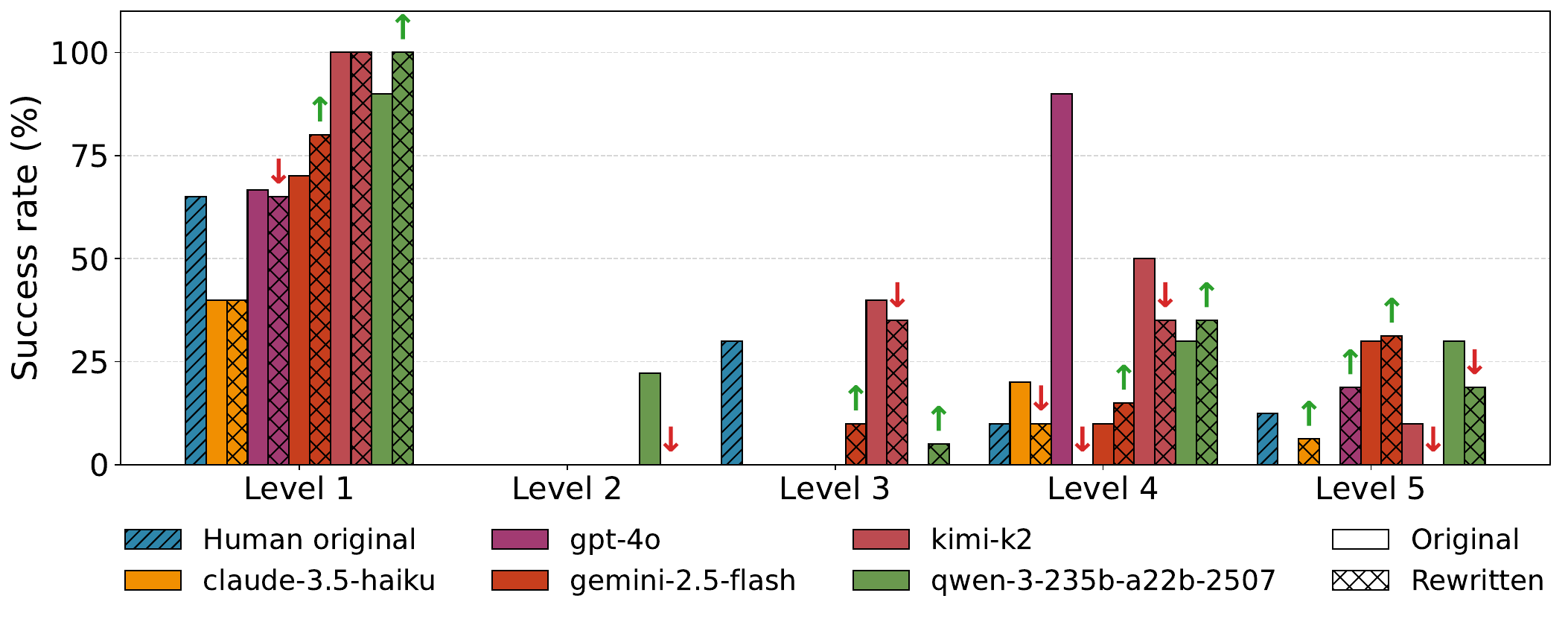} \caption{ Integrated success rates under a Claude 4.5 Sonnet judge.
Rewriting often improves over human-only and can sometimes improve over LLM-only, but it depends on the level and writer. }
\label{fig:hybrid_advantage_claude}
\end{figure}
\begin{figure}[htbp]
\centering \includegraphics[width=\linewidth]{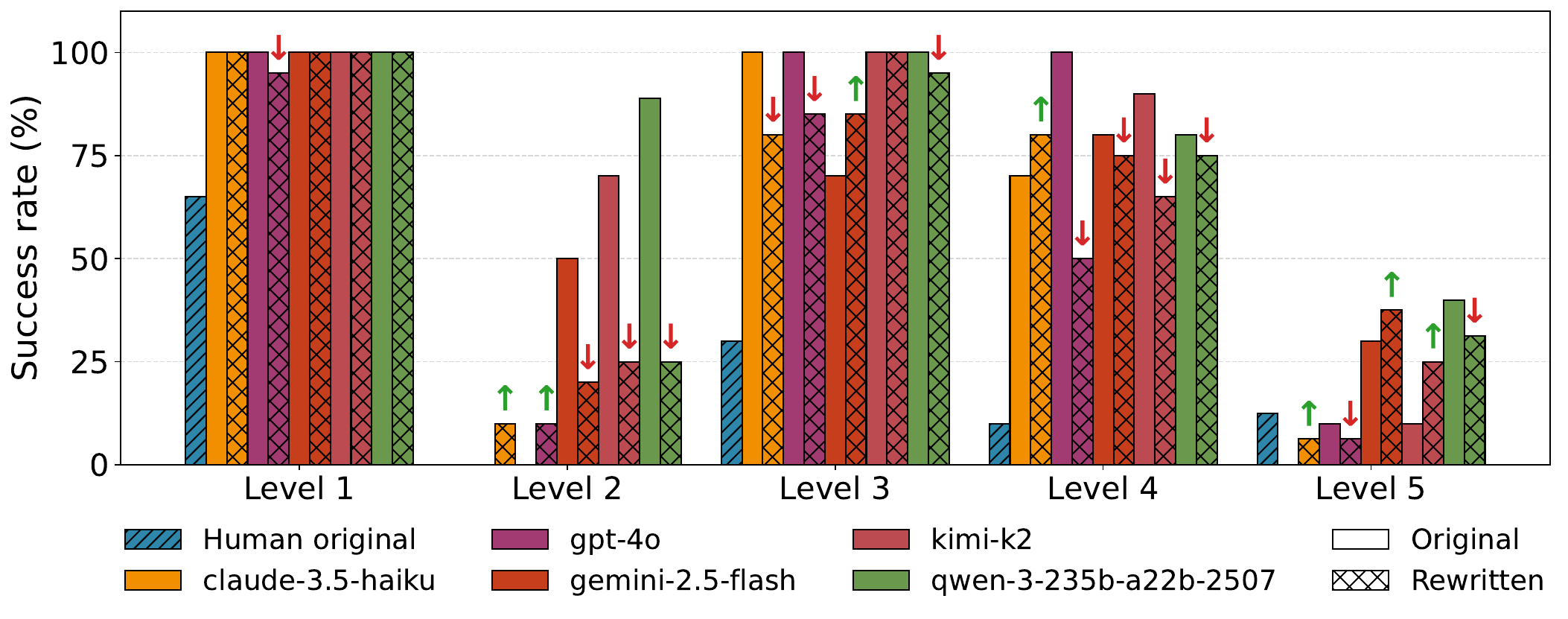} \caption{ Integrated success rates under a Grok 4 Fast judge.
Hybrid gains vary by scenario level and writer model. }
\label{fig:hybrid_advantage_grok}
\end{figure}

Conversely, when the LLM-only baseline is low, rewriting can yield large absolute improvements.
Under the GPT-5.2 judge (Figure~ \ref{fig:hybrid_advantage_gpt52}), Claude 3.5 Haiku has only 10\% pass rate in LLM-only at level 5, but rises to 62.5\% when rewriting a human draft.
Similarly, under the same judge, Gemini 2.5 Flash improves from 10\% to 40\% at level 4, and Qwen 3 235B improves from 30\% to 55\% at level 3.
Under the Claude 4.5 Sonnet judge (Figure~\ref{fig:hybrid_advantage_claude}), GPT-4o has 0\% LLM-only at level 5 but reaches 18.75\% after rewriting.
Under the Grok 4 Fast judge (Figure~\ref{fig:hybrid_advantage_grok}), Kimi K2 rises from 10\% to 25\% at level 5.
These cases illustrate the headroom effect: hybrid is most likely to look strong when the writer's LLM-only baseline is far from saturated.

Finally, the hardest level (level 5) is where the hybrid effects are the most judge-sensitive.
Under GPT-4o as judge, the hybrid advantage is weakest on level 5 (Figure~\ref{fig:hybrid_advantage}); in contrast, under GPT-5.2 and Grok 4 Fast there are clearer regimes where rewriting helps on level 5 (Figures~\ref{fig:hybrid_advantage_gpt52} and \ref{fig:hybrid_advantage_grok}).
We view this as evidence that the benefit of rewriting depends not only on the writer model but also on the evaluation criteria implicit in the judge and on how much improvement is available over the writer's own baseline in that scenario.

\clearpage

\clearpage
\section{Email tone analysis}
\label{appendix:tone_analysis}

This appendix documents the procedure used to label emails along two tone dimensions---empathy and formality---used in Figure~\ref{fig:emp_form_analysis}.
Our goal is to characterize differences in writing style between human and LLM emails, and to understand how LLM rewriting changes the tone of a human draft.

\paragraph{Label definitions.}
\textit{Empathy} measures whether an email acknowledges the recipient's perspective and tries to help with their concern.
We instruct labelers to use a 1--7 Likert scale anchored as: (1) cold, no compromise or concern; (4) neutral or mixed tone; (7) warm, accommodating, clearly empathetic.
\textit{Formality} measures whether the tone is casual/blunt versus polished and corporate.
We instruct labelers to use a 1--7 Likert scale anchored as: (1) highly informal or blunt; (4) neutral; (7) polished corporate/professional.

\paragraph{LLM labeler, prompting, and decoding settings.}
We score emails with an LLM-as-a-judge prompt that requests paragraph-level ratings and a strict JSON response.
For the plots in this version of the paper we used \texttt{google/gemini-3-flash-preview} via OpenRouter with temperature 0.0.
To reduce variance, we use deterministic decoding (temperature 0).

\paragraph{System prompt.} \mbox{} \\
\begin{promptbox}
You are an expert communication coach who scores workplace emails.
Use the provided Likert scales strictly.
\end{promptbox}

\paragraph{User prompt.}
The user message contains (i) the Likert definitions, (ii) scenario context, and (iii) the email text, and instructs the labeler how to segment paragraphs and index them:
\begin{promptbox}
Rate the following email independently on two 1-7 scales.

Empathy scale: 1 = cold, no compromise or concern. 4 = neutral or mixed tone. 7 = warm, accommodating, clearly empathetic.

Formality scale: 1 = highly informal or blunt. 4 = neutral, not too formal or informal. 7 = polished corporate/professional.

Scenario context:
$<$scenario prompt text for the corresponding level$>$

Evaluate each paragraph separately.
Paragraphs are separated by blank lines; do not treat a standalone greeting or the sign-off/signature as paragraphs for rating.
Start paragraph indexing at the first body paragraph after the greeting.

Email:
$<$email text$>$

Respond with JSON only: \{ "paragraph\_ratings": [ \{ "paragraph\_index": 1, "empathy\_score": $<$integer 1--7$>$, "empathy\_rationale": "$<$brief reason. 1 sentence max.$>$", "formality\_score": $<$integer 1--7$>$, "formality\_rationale": "$<$brief reason. 1 sentence max.$>$" \}, ... ] \}
\end{promptbox}

\paragraph{Paragraph and turn handling.}
We ask the labeler to treat paragraphs as blocks separated by blank lines and to ignore standalone greetings and sign-offs/signatures.
The labeler outputs one rating per body paragraph, with indices starting from the first body paragraph after the greeting.

Level 4 interactions can be multi-turn.
To make tone comparable across levels, we score each turn separately.
Turns are explicitly tagged with speaker markers, e.g., \texttt{Turn 1 - Alex:} (player) and \texttt{Turn 1 - Adam:} (recipient).
We extract only the player turns and drop recipient turns before labeling.
Operationally, this means a single level-4 thread can contribute multiple tone-labeled records (one per player turn).
When we plot level-4 points, each turn is treated as its own email instance in empathy--formality space.

For each email (or turn), the labeler outputs paragraph-level scores and brief rationales in the JSON format shown above.
For downstream analysis and plotting, we compute per-email averages by taking the arithmetic mean over paragraphs: \(\overline{e} = \frac{1}{P}\sum_{p=1}^{P} e_p\) and \(\overline{f} = \frac{1}{P}\sum_{p=1}^{P} f_p\).

\paragraph{Inter-annotator agreement.}
To validate the reliability of the tone labels, we measure inter-annotator agreement (IAA) between human annotators and LLM annotators on the common subset of emails rated by all annotators.
Agreement is computed separately for empathy and formality using two complementary metrics.

The first metric is \textbf{Krippendorff's \(\alpha\)}, a chance-corrected agreement coefficient defined on an interval scale~\citep{krippendorff2011computing}.
For \(n\) emails and \(C\) annotators, let \(v_{ci}\) denote the per-email average score assigned by annotator \(c\) to email \(i\).
The coefficient is \(\alpha = 1 - \frac{D_o}{D_e}\), where:
\begin{align}
D_o &= \frac{1}{n} \sum_{i=1}^{n} \frac{1}{C(C-1)} \sum_{c \neq c'} (v_{ci} - v_{c'i})^2, \\
D_e &= \frac{1}{nC(nC-1)} \sum_{(i,c) \neq (i',c')} (v_{ci} - v_{c'i'})^2.
\end{align}
\(D_o\) is the mean squared difference between all pairs of annotators \emph{on the same email}, averaging over emails.
\(D_e\) is the mean squared difference over all pairs of email scores (regardless of annotator), and serves as a baseline for chance-level disagreement.
We use the interval distance function \(d(a, b) = (a - b)^2\).
\(\alpha = 1\) indicates perfect agreement, \(\alpha = 0\) indicates agreement at chance level, and \(\alpha < 0\) indicates systematic disagreement worse than chance.

The second metric is the \textbf{average pairwise Spearman correlation} \(\bar{\rho}\).
For each pair of annotators \((i, j)\), we compute the Spearman rank correlation \(\rho_{ij}\) between their per-email average scores across all commonly annotated emails.
We then average over all \(\binom{C}{2}\) pairs: \(\bar{\rho} = \frac{2}{C(C-1)}\sum_{c < c'} \rho_{cc'}\).
Spearman correlation captures whether annotators produce consistent \emph{rankings} of emails, regardless of any absolute difference in scores.
Together, Krippendorff's \(\alpha\) and \(\bar{\rho}\) provide complementary views: the former measures absolute agreement on scores, while the latter measures agreement on the relative ordering of emails.

Table~\ref{tab:iaa} reports the agreement between the three annotators (two human and one LLM) on the common set of 4 emails.
We see good agreement between the annotators (Krippendorff's \(\alpha\) 0.620 and 0.429 for empathy and formality, respectively) and good agreement on the relative ordering of emails (Spearman correlation 0.600 and 0.821 for empathy and formality, respectively).

\begin{table}[h]
\centering
\begin{tabular}{lcc}
\toprule
\textbf{Dimension} & \textbf{Krippendorff's \(\alpha\)} & \textbf{Avg.\ Pairwise \(\bar{\rho}\)} \\
\midrule
Empathy   & 0.595 & 0.676 \\
Formality & 0.578 & 0.731 \\
\bottomrule
\end{tabular}
\caption{Inter-annotator agreement across three annotators (two humans, one LLM) on 4 commonly annotated emails.}
\label{tab:iaa}
\end{table}

\paragraph{Examples of rewrite mechanics.}
Table~\ref{tab:rewrite_mechanics_examples} provides qualitative examples of how
LLM rewrites change specific sentences in a human draft to increase empathy
and/or formality.

\begin{table*}[t]
    \centering
    \small
    \setlength{\tabcolsep}{3pt}
    \renewcommand{\arraystretch}{1.15}
    \begin{tabularx}{\textwidth}{@{}cXXX@{}}
    \toprule
    \textbf{Scen.} & \textbf{Human sentence(s)} & \textbf{LLM rewrite sentence(s)} & \textbf{Mechanics / tone shift} \\
    \midrule
    1 & {\ttfamily Unfortunately our finance department has been using AI-generated advice for taxes, and we got hit HARD by last year's audit. Whoops!} & {\ttfamily Welcome aboard—we’re so excited to have you on the team! I wanted to take a moment to address your recent question about having a private office.} & \textbf{GPT-4o}; $\Delta$ Empathy: +4.6; $\Delta$ Formality: +4.2; Empathy: The tone is warm and welcoming, showing immediate enthusiasm for the new hire; Formality: The language is professional but uses a friendly, conversational opening \\
    2 & {\ttfamily if you two don't work together, you're both fired. No punchlines, no slogans, no mission focus -- pure visual campaign.} & {\ttfamily I hope you're both doing well and finding moments to recharge amid the project push. I know the deadline for the AI Agent Optimizer App ad is coming up on March 20, and I appreciate the dedication you’ve both shown so far.} & \textbf{Qwen 3}; $\Delta$ Empathy: +5.0; $\Delta$ Formality: +4.2; Empathy: The opening is warm, acknowledges their hard work, and expresses genuine appreciation for their dedication; Formality: The tone is professional and polished but maintains a friendly, accessible corporate style \\
    3 & {\ttfamily Dear Dave, it has come to our attention that you do not agree with our back to the office policy . Unfortunately, we are a very rigid company and once we come up with policies, we adhere to them without taking into account any special circumstance...} & {\ttfamily Congratulations on the arrival of your newborn son! I know balancing work and family can be challenging, which is why I want to discuss our upcoming workplace changes thoughtfully.} & \textbf{Claude 3.5 Haiku}; $\Delta$ Empathy: +5.0; $\Delta$ Formality: +1.8; Empathy: The paragraph opens with a warm personal congratulation and explicitly acknowledges the difficulty of balancing work and family; Formality: The tone is professional and polished while remaining accessible and supportive \\
    4 & {\ttfamily TURN 5: adam just tell me what you want man.} & {\ttfamily I’ve reviewed your feedback and the recent adjustments made to address your concerns, and I want to ensure we’re fully supporting you—not just in resolving immediate issues, but in creating a work environment where you can continue to thrive.} & \textbf{Qwen 3}; $\Delta$ Empathy: +4.9; $\Delta$ Formality: +5.3; Empathy: The tone is warm and focuses on supporting the employee's ability to thrive rather than just fixing problems; Formality: The language is polished, professional, and uses sophisticated corporate phrasing \\
    5 & {\ttfamily Core Issues: * poor work ethic} & {\ttfamily I hope this email finds you well. First, I want to acknowledge the effort and time you’ve put into your role at NeuroGrid in the past.} & \textbf{GPT-4o}; $\Delta$ Empathy: +3.5; $\Delta$ Formality: +4.2; Empathy: The tone is warm and validating, acknowledging the difficulty of career growth; Formality: Uses standard professional opening and polished phrasing \\
    \bottomrule
    \end{tabularx}
    \caption{Qualitative examples of how LLM rewrites shift a human draft's tone in empathy--formality space (Figure~\ref{fig:human_llm_vectors}). Each row pairs a human email with an LLM rewrite of the same index and scenario.}
    \label{tab:rewrite_mechanics_examples}
\end{table*}

\paragraph{Examples of low-empathy, low-formality human emails.}
To illustrate what ``low empathy, low formality'' looks like, below we show one representative human example from each level (1--5).
In levels 1, 2, 4, and 5 the example falls in the low-empathy, low-formality quadrant (avg empathy \(<4\), avg formality \(<4\)).
In level 3, the lowest-empathy example in our sample has neutral formality (avg formality \(=4.0\)).

\begin{promptbox}
\textbf{Level 1 (avg empathy 1.0, avg formality 1.0).}\par {\ttfamily don't worry about it bro, thats a you problem.}
\end{promptbox}

\begin{promptbox}
\textbf{Level 2 (avg empathy 1.0, avg formality 1.0).}\par {\ttfamily if you two don't work together, you're both fired.
No punchlines, no slogans, no mission focus -- pure visual campaign.
Get to it}
\end{promptbox}

\begin{promptbox}
\textbf{Level 3 (avg empathy 1.0, avg formality 4.0).}\par {\ttfamily Dear Dave, it has come to our attention that you do not agree with our back to the office policy.
Unfortunately, we are a very rigid company and once we come up with policies, we adhere to them without taking into account any special circumstances, like the ones that you clearly have in terms of childcare and having a spouse who is out of the house.
We hope you understand our lack of flexibility in this regard, and we wish you the best.}
\end{promptbox}

\begin{promptbox}
\textbf{Level 4 (turn; avg empathy 1.0, avg formality 1.0).}\par {\ttfamily TURN 4:\\ Brittany, if you are reading this email, please stop.
You're useless.\\ \\ Adam: STOP RESPONDING.
OR WE WILL TAKE LEGAL ACTION}
\end{promptbox}

\begin{promptbox}
\textbf{Level 5 (avg empathy 1.0, avg formality 1.0).}\par {\ttfamily jake if you fuck up again ill kill you}
\end{promptbox}


\end{document}